\documentclass[sigconf, screen]{acmart}

\PassOptionsToPackage{dvipsnames}{xcolor}
\PassOptionsToPackage{colorlinks=true,linkcolor=brown,urlcolor=blue,breaklinks=True,citecolor=RubineRed}{xcolor}

\usepackage{enumitem}
\setlist[itemize]{noitemsep, topsep=0pt}

\usepackage[most]{tcolorbox}
\usepackage{adjustbox}

\usepackage{multirow}
\usepackage{booktabs,subcaption,dcolumn}
\usepackage{tikz}
\usepackage{enumitem}

\usepackage{MnSymbol}

\usepackage{pifont}
\usepackage{svg}
\usepackage{soul}
\usepackage{amsmath}
\usepackage{dirtytalk}
\usepackage{subcaption}
\usepackage{graphicx}
\usepackage{setspace}
\usepackage{caption}
\usepackage{booktabs} 
\usepackage[titletoc,toc,title]{appendix}
\usepackage{svg}
\usepackage{amsfonts}
\usepackage{xurl} 

\usepackage{diagbox}

\usepackage{framed}

\usepackage{listings}
\usepackage{xcolor}

\newcounter{emailbox}                  

\lstdefinelanguage{json}{
  morestring=[b]",
  morecomment=[l]{//},
  morekeywords={true,false,null},
  sensitive=false,
}

\usepackage{amsmath}
\usepackage{amsfonts} 

\usepackage{colortbl}

\newcolumntype{?}{!{\vrule width 1.5pt}}

\newcommand\smamath[1]{{\small $#1$}}

\newcommand\revision[1]{%
  \bgroup
  \hskip0pt\color{blue!80!black}%
  #1%
  \egroup
}

\newcommand\purple[1]{%
  \bgroup
  \hskip0pt\color{purple!50!blue}%
  #1%
  \egroup
}

\newcommand\todo[1]{%
  \bgroup
  \hskip0pt\color{red!80!black}%
  #1%
  \egroup
}

\newcommand{\ds}[1][\small]{{#1E-PhishLLM}}
\newcommand{\fmw}[1][\small]{{#1E-PhishGEN}}

\newtcolorbox{cooltextbox}[1][]{%
    colback=black!5,
    colframe=black!5,
    notitle,
    sharp corners,
    borderline west={0pt}{0pt}{red!80!black},
    enhanced,
    breakable,
    left=0pt,
    right=0pt,
    top=0pt,
    bottom=0pt
    }

\newtcolorbox{emailBox}[1]{%
  colback=gray!5!white, 
  colframe=gray!50!white, 
  boxrule=0.5pt, 
  sharp corners, 
  title={\textbf{[Subject] #1}}, 
  fonttitle=\small\sffamily, 
  after lower={}, 
  nobeforeafter, 
  breakable, 
      left=2pt,
    right=2pt,
    top=0pt,
    bottom=0pt,
    before upper=\refstepcounter{emailbox}
}

\definecolor{pastegreen}{rgb}{0.61, 0.87, 0.53}

\newcounter{promptcounter} 

\newtcolorbox[use counter=promptcounter, number within=none]{prompt}[2][]{colback=pastegreen!10, colframe=pastegreen!80, 
    coltitle=black, fonttitle=\bfseries, title=Prompt~\thepromptcounter: #2, #1, sharp corners, boxrule=0.8mm, fontupper=\footnotesize,left=2pt,
    right=2pt,
    top=0pt,
    bottom=0pt}
\renewcommand{\thepromptcounter}{\arabic{promptcounter}}

\AtBeginDocument{%
  }

\copyrightyear{2025}
\acmYear{2025}
\setcopyright{acmlicensed}
\acmConference[AISec '25]{Proceedings of the 2025 Workshop on Artificial Intelligence and Security}{October 13--17, 2025}{Taipei, Taiwan}
\acmBooktitle{Proceedings of the 2025 Workshop on Artificial Intelligence and Security (AISec '25), October 13--17, 2025, Taipei, Taiwan}
\acmDOI{10.1145/3733799.3762967}
\acmISBN{979-8-4007-1895-3/2025/10}

\begin{document}


\title[E-PhishGen: Unlocking Novel Research in Phishing Email Detection]{E-PhishGen: Unlocking Novel Research\\in Phishing Email Detection}

\author{Luca Pajola}
\authornote{Luca Pajola is the first author and should be contacted for correspondence about this work. Eugenio Caripoti and Stefan Banzer have both contributed substantially to this work. SpritzMatter S.R.L. is a spinoff of the University of Padua.}
\orcid{1234-5678-9012}
\affiliation{%
  \institution{SpritzMatter S. R. L.}
  \city{Padua}
  \country{Italy}
}
\affiliation{%
  \institution{University of Padua}
  \city{Padua}
  \country{Italy}
}
\email{luca.pajola@spritzmatter.com}

\author{Eugenio Caripoti}
\authornotemark[1]
\orcid{0009-0007-3753-1859}
\affiliation{%
  \institution{SpritzMatter S. R. L.}
  \city{Padua}
  \country{Italy}
}
\affiliation{%
  \institution{University of Padua}
  \city{Padua}
  \country{Italy}
}
\email{eugenio.caripoti@spritzmatter.com}

\author{Stefan Banzer}
\authornotemark[1]
\orcid{0009-0007-4727-6468}
\affiliation{%
  \institution{University of Liechtenstein}
  \city{Vaduz}
  \country{Liechtenstein}
}
\email{stefan.banzer@uni.li}

\author{Simeone Pizzi}
\authornotemark[1]
\orcid{0009-0007-6719-0813}
\affiliation{%
  \institution{SpritzMatter S.R.L.}
  \city{Padua}
  \country{Italy}
}
\email{simeone.pizzi@spritzmatter.com}

\author{Mauro Conti}
\orcid{0000-0002-3612-1934}
\affiliation{%
  \institution{University of Padua}
  \city{Padua}
  \country{Italy}}
\affiliation{%
  \institution{Orebro University}
  \city{Orebro}
  \country{Sweden}}
\email{mauro.conti@unipd.it}

\author{Giovanni Apruzzese}
\orcid{0000-0002-6890-9611}
\affiliation{%
  \institution{University of Liechtenstein}
  \city{Vaduz}
  \country{Liechtenstein}}
\affiliation{%
  \institution{University of Reykjavik}
  \city{Reykjavik}
  \country{Iceland}}
\email{giovanni.apruzzese@uni.li}


\begin{abstract}

Every day, our inboxes are flooded with unsolicited emails, ranging between annoying spam to more subtle phishing scams. Unfortunately, despite abundant prior efforts proposing solutions achieving near-perfect accuracy, the reality is that countering malicious emails still remains an unsolved dilemma.

This ``open problem'' paper carries out a critical assessment of scientific works in the context of phishing email detection. First, we focus on the \textit{benchmark datasets} that have been used to assess the methods proposed in research. We find that most prior work relied on datasets containing emails that---we argue---are not representative of current trends, and mostly encompass the English language. Based on this finding, we then re-implement and re-assess a variety of \textit{detection methods reliant on machine learning}~(ML), including large-language models (LLM), and release all of our codebase---an (unfortunately) uncommon practice in related research. We show that most such methods achieve near-perfect performance when trained and tested on the same dataset---a result which intrinsically hinders development (how can future research outperform methods that are already near perfect?). To foster the creation of ``more challenging benchmarks'' that reflect current phishing trends, we propose \fmw{}, an LLM-based (and privacy-savvy) framework to generate novel phishing-email datasets. We use our \fmw{} to create \ds{}, a novel phishing-email detection dataset containing 16616 emails in three languages. We use \ds{} to test the detectors we considered, showing a much lower performance than that achieved on existing benchmarks---indicating a larger room for improvement. We also validate the quality of \ds{} with a user study (n=30). To sum up, we show that phishing email detection is still an open problem---and provide the means to tackle such a problem by future research.

\end{abstract}


\keywords{benchmark, dataset, large language models, email, spam, detection}

\maketitle

\section{Introduction}
\label{sec:introduction}

\noindent
Hardly a day goes by without hearing of yet another company being targeted by a phishing attack~\cite{hoxhunt2025report,apwg2024report}. In particular, phishing \textit{emails} still represent one of the most common vectors to penetrate an organization and carry out any sort of cyberattack---ranging from gathering login credentials, installing malware, or stealing industrial secrets~\cite{hoxhunt2025report}. Simply put, the reality is that the battle against phishing-email attacks is never in favor of system defenders.

And yet, by turning the attention at prior research on \textit{phishing email detection}, the conclusions of a large number of papers seem to align: phishing emails can be detected with near-perfect performance by using diverse techniques within the machine-learning (ML) domain. For instance, Bountakas and Xenakis~\cite{bountakas2023helphed} claim a random forest classifier achieves 98.6\% accuracy; Doshi et al.~\cite{doshi2023comprehensive} achieve 98.4\% accuracy with a logistic regression model; whereas Atawneh and Aljehani
~\cite{atawneh2023phishing} use a BERT model obtaining 99.6\% accuracy; large-language models (LLM) have also been used, achieving accuracy above 95\% ~\cite{beydemir2024dynamically,xue2025multiphishguard}. Put simply, the scenario portrayed in research depicts a remarkably different, and almost contradictory, reality than what is faced by real-world companies.

In this work, we seek to 
{\small \textit{(i)}}~provide evidence that the detection of phishing emails is an open problem \textit{in research},
{\small \textit{(ii)}}~identify potential root causes, and 
{\small \textit{(iii)}}~devise the means to address such a problem by future research. Let us explain how we pursue our goals.

\vspace{2mm}

\noindent
\textsc{\textbf{Research Questions and Findings.}} We begin our quest with Section~§\ref{sec:literature}, which revolves around a broad, and our first, research question (RQ1): \textit{``\purple{what benchmark datasets are used in related literature to assess previously-proposed phishing email detectors?}''} To answer RQ1, we review related literature, and find that most prior works relied on a subset of eight datasets, which share three problems: {\small \textit{(a)}}~they mostly include English emails---neglecting other languages, such as German or Italian, in which phishing emails can be written; {\small \textit{(b)}}~they often mix ``phishing'' with ``spam'' emails---the latter not necessarily posing a security/privacy risk; {\small \textit{(c)}}~they predominantly include emails collected before 2010---which are unlikely to reflect current phishing trends. To provide a clear case, we report in Email~\ref{email:spamassassin} a sample contained in the SpamAssassin dataset~\cite{spamassassin} (used, e.g., to test recent approaches proposed in~\cite{doshi2023comprehensive,atawneh2023phishing,xue2025multiphishguard}). Altogether, these findings show the limited scope of existing benchmark datasets, i.e., a crucial component in an experimental evaluation.

\vspace{2mm}
\begin{emailBox}{Help!}
{\scriptsize
You have been specially selected to qualify for the following:

\textbf{Premium Vacation Package and Pentium PC Giveaway}

To review the details, please click on the link below using the confirmation number:

\url{http://www.1chn.net/wintrip} 

Confirmation Number: \textbf{Lh340}

Please confirm your entry within 24 hours of receiving this confirmation.

Wishing you a fun-filled vacation!

If you have any additional questions or cannot connect to the site, do not hesitate to contact 

me: 

\texttt{vacation@btamail.net.cn}}
\end{emailBox}
\vspace{-1mm}
{\noindent\small\textbf{Email~\ref{email:spamassassin}.} An email in the popular dataset SpamAssassin~\cite{spamassassin} (from 2005).}
\label{email:spamassassin}
\vspace{3mm}

Then, in Section~\ref{sec:reassessment}, we tackle our second research question (RQ2): ``\textit{\purple{what performance do existing detectors achieve on some previously-used benchmark datasets?}}'' Indeed, we found that few prior works released their codebase, and even though the results of prior works aligned, it was difficult to pinpoint a clear baseline. Hence, to facilitate future research, we carry out a reproducible and statistically-validated \textit{reassessment of previously-proposed phishing email detectors} on the identified benchmark datasets. We consider detection techniques reliant on feature-engineering (e.g., TF-IDF) as well as those based on ``feature-agnostic'' techniques such as BERT, and we also consider detectors that leverage pretrained large-language models (LLMs) in a zero-shot fashion. Our results confirm that ML methods trained and tested on the same dataset achieve near-perfect performance---which is not a very encouraging result, because it indicates that existing benchmarks are not adequate to measure the ``improvement'' that a given method may have over existing ones. As a potential workaround, we also carry out a cross-evaluation~\cite{apruzzese2022cross} by testing our considered detection methods on different datasets. Our results show an (expected) drop in performance---suggesting that a similar (but not very common, in this domain) practice may be more appropriate to gauge the quality of novel detectors.

Next, in Section~\ref{sec:framework}, we turn the attention to our third research question (RQ3): ``\textit{\purple{what is a way to overcome the shortcomings of existing datasets---without raising privacy concerns?}}'' Indeed even by mixing existing datasets, their limited scope (e.g., few languages, old emails) does not enable one to assess the real-world effectiveness of existing detectors. A potential solution is collecting emails from ``our inboxes'', but such a practice may raise privacy concerns~\cite{kyaw2024systematic}, especially given that a benchmark dataset must be publicly released. Therefore, we {\small \textit{(a)}}~propose a framework, \fmw{}, that enables researchers to ~create custom datasets of phishing emails that conform to user-provided characteristics; and {\small \textit{(b)}}~use our proposed framework to create \ds{}, a novel dataset of 16616 emails (equally split between benign and phishing) encompassing different languages and ``personal profiles'', thereby overcoming the shortcomings of existing datasets.

Finally, in Section~\ref{sec:validation}, we focus on our fourth and fifth research questions, specifically: ``\purple{\textit{what performance do previous methods achieve on \ds{}?}}'' (RQ4) and ``\purple{\textit{does \ds{} contain phishing emails of a higher quality than those included in previously-proposed datasets?}}'' (RQ5). Answering these RQs serves to {\small \textit{(i)}}~provide a well-founded baseline for future work that seeks to use \ds{} for benchmarking purposes, but also {\small \textit{(ii)}}~validate our claim that \ds{} overcomes commonly used benchmark datasets. For RQ4, we test the methods considered in RQ2 on the new dataset, whereas for RQ5 we carry out a user study with experts (n=30).

We discuss limitations and ethics (including the potential ``dual use'' of \fmw{}) in Section~\ref{sec:discussion}. We release all of our resources~\cite{repository2}.

\vspace{2mm}

\noindent
\textsc{\textbf{Technical Contributions.}}
Despite being primarily an ``open problem'' paper, we do make three technical contributions to the domain of applied ML for cybersecurity. Namely:
\begin{itemize}[leftmargin=*]
    \item We propose \fmw{}, an original framework to generate targeted phishing emails tailored for user-specific profiles.
    \item We provide \ds{}, a novel dataset of phishing emails---generated via \fmw{}---that are qualitatively better than those contained in existing datasets (validated via an user study).
    \item We reassess previously-proposed and ML-based phishing email detection methods on existing datasets as well as on \ds{}.
\end{itemize}
In summary, we provide the tools to ``unlock'' the relatively stagnant (in our opinion) state of research on phishing email detection. 
\section{Literature Review and Motivation}
\label{sec:literature}

\noindent
We briefly introduce the problem of phishing email detection~(§\ref{ssec:background}), then we address RQ1 (§\ref{ssec:datasets}), and finally discuss related work (§\ref{ssec:related}). 
Importantly, we focus on phishing \textit{emails}: other phishing vectors (e.g., websites~\cite{abdelnabi2020visualphishnet}, SMS~\cite{timko2024smishing}, voice~\cite{gupta2015phoneypot}) are outside our scope.

\subsection{Background: Phishing Emails}
\label{ssec:background}

\noindent
Phishing is a form of cyberattack that has been plaguing internet users by over twenty years~\cite{dhamija2006phishing,baki2023sixteen}. The overarching principle of phishing is to leverage social-engineering techniques that induce a given victim to give the attacker what they want. For instance, an employee may receive an email from their supervisor (whose real email address has been compromised or spoofed) asking to open an attachment (which may contain, e.g., ransomware); or they may receive a fake email from their security department asking to change their login credentials, pointing to a malicious link~\cite{weinz2025impact}. 

\textbf{Phishing Emails (in the real world).}
Every year, new reports from renown sources (e.g., Proofpoint~\cite{proofpoint2024phish}, APWG~\cite{apwg2024report}, Cisco Talos~\cite{talos2025trends}, or even the FBI~\cite{fbi2024icr}) reveal that companies of all sizes are constantly, and increasingly, targeted by phishing threats---especially after the recent release of LLMs~\cite{weinz2025impact,proofpoint2024phish}. Among the most common recommendations against such a (almost never-ending) problem is phishing training/education~\cite{lain2024content,weinz2025impact,lain2022phishing}.
However, such defensive practices are still not widespread~\cite{proofpoint2024phish}, and even ``trained'' employees may occasionally fall victim to a phishing scam, potentially compromising their entire organization~\cite{weinz2025impact,lain2022phishing}. Therefore, there is a need to develop automated detection techniques that can prevent end-users from receiving (or opening) a phishing email in the first place. In practice, this is typically done via hard-coded mechanisms such as blocking/flagging emails from unknown senders, or that contain attachments with a known malicious signature, or that include specific terms/URLs in the text~\cite{biggio2018wild,weinz2025impact,fratantonio2025magika}. However, all such methods can be easily circumvented by real-world attackers (after all, we do receive a lot of phishing emails!) denoting that currently-deployed detectors/filters present huge margins for improvement.

\textbf{Phishing Email Detection (in research).} A variety of approaches, typically reliant on machine learning (ML), have been proposed in research to mitigate the problem of phishing emails. At a high-level, such methods seek to overcome the intrinsic limitation of ``static'' filters, which can only work against phishing emails that match pre-defined rules (e.g., inclusion of an URL reported in a blocklist). Development of ML-based detectors typically requires three steps~\cite{apruzzese2022sok}: First, a representative dataset of phishing and benign emails must be collected; potentially, the dataset must be pre-processed to enable the application of ML methods. Second, a given ML model must be trained on the (pre-processed) dataset. Third, the ML model must be tested (on different data): if the performance of such ML model meets certain requirements (e.g., low false positives with high true positives) then it can be deployed in operational environments. Broadly, we can distinguish three categories of ML-based phishing-email detectors:
\begin{itemize}[leftmargin=*]
    \item \textit{Feature-based.} These detectors analyse manually-engineered features (e.g., occurrence of certain words in the email's text). Therefore, such methods require a feature-extraction process that turns the original email into a feature-vector, representing the sample provided as input to the ML model. For instance, Doshi et al.~\cite{doshi2023comprehensive} apply TF-IDF on the email's text before sending the sample to the ML-based classifier; their evaluation revealed that the best-performing model relied on the logistic regression (LR) algorithm.
    \item \textit{Feature-agnostic.} These detectors send the whole email directly to the ML model---and hence do not require any pre-processing step. However, it is necessary to carry out some training/fine-tuning of the model. For instance, Fang et al.~\cite{fang2019phishing} use a convolutional neural network that receives the output of the Word2Vec algorithm to perform their detection; whereas Lee et al.~\cite{lee2021d} (as well as~\cite{atawneh2023phishing}) use language models, such as BERT.
    \item \textit{LLM-based.} These detectors can work without any preprocessing or fine-tuning. The intuition is to leverage ``large'' language models to autonomously carry out the detection task, potentially by issuing one/few prompts---as done in, e.g.,~\cite{beydemir2024dynamically,xue2025multiphishguard}.
\end{itemize}
Of course, a detector can also rely on a combination of the aforementioned categories, e.g., by combining different detectors in an ensemble, pipeline, or high-level architecture (as done, e.g., in~\cite{lee2021d}).

\subsection{Datasets for Phishing Email Detection {\normalsize [RQ1]}}
\label{ssec:datasets}
\noindent
In research, whenever a paper proposes any given solution, such a paper should demonstrate that the proposed method ``outperforms the state of the art'' (or ``achieves state of the art performance''). Such a demonstration requires an evaluation---which necessitates a dataset. As we wrote, the evaluations of previously-proposed methods for phishing email detection typically show near-perfect performance---a finding confirmed by recent literature reviews~\cite{das2019sok,alhuzali2025depth,salloum2022systematic,champa2024curated,kyaw2024systematic,thakur2023systematic}. \textit{If prior research shows such a stunning performance, then why is it that phishing emails keep flooding our inboxes?} Such a dilemma led us to scrutinize the datasets used to test the methods of prior work---i.e., RQ1. Let us explain how we ``motivate'' our paper.

\textbf{Methodology.} We acknowledge that prior reviews have tackled a similar question (e.g.,~\cite{das2019sok,alhuzali2025depth,salloum2022systematic,champa2024curated,kyaw2024systematic,thakur2023systematic}); moreover, investigating all literature on phishing-email detection is unfeasible. To provide a meaningful, and humanly-possible, answer to RQ1 we carry out a semi-systematic literature review~\cite{zunder2021semi}, proceeding as follows. 
\textbf{(1)}~We search well-known repositories (Google Scholar, IEEE Xplore, Tailor\&Francis) with queries ``[LLM/large language model/] phishing email [detection/classification]'', and we integrate our results with papers accepted to top-conferences (WWW, S\&P/EuroS\&P, CCS/AsiaCCS, NDSS, USENIX SEC, ACSAC, IMC, WDSM, CHI) in 2014--2024 that have ``phish'' in the title (as also done in~\cite{weinz2025impact}). This led to 562 papers (for Google Scholar, we only considered the first 100 results returned with each query). 
\textbf{(2)}~We filtered our papers by looking at their metadata. For instance, we used the abstract to determine if the paper was truly about ``phishing email detection'' (and not, e.g., phishing ``website'' detection), and we generally excluded unpublished articles. This led us to a set of 100 papers. 
\textbf{(3)}~We qualitatively analysed the full-text of these papers. Specifically, attention was put on the following aspects: {\small \textit{(i)}}~what datasets are used?
{\small \textit{(ii)}}~what detection methods are proposed?
{\small \textit{(iii)}}~is the source code available? Such a qualitative analysis has been carried out by four researchers, who worked independently and discussed their findings in meetings (as also done in~\cite{apruzzese2023real}).

\textbf{Qualitative Findings.} 
We report the results of our dataset-centered analysis (detection approaches are covered in §\ref{sec:reassessment}).
\begin{itemize}[leftmargin=*]
    \item \textit{Overly-used datasets.} The vast majority of our analysed papers relies on the same datasets to test their methods: SpamAssassin~\cite{spamassassin}, Enron~\cite{klimt2004enron}, SpamBase~\cite{spambase_94}, Nazario~\cite{nazario}, LingSpam~\cite{sakkis2003memory}, NigerianFraud/Clair~\cite{clair}, TREC~\cite{trec}, CEAS~\cite{ceas}. While reliance on the same dataset is not a bad practice per-se (after all, it enables comparisons across different methods), it becomes such when one considers that such datasets (aside from Nazario) have emails collected mostly before 2010. Such ``old'' data clearly does not reflect the most recent phishing trends.\footnote{We argue that 99\% accuracy on detecting phishing emails exchanged \textit{in 2005} cannot be used to claim that a method is an ``efficient detector of phishing emails'' after 2020.} Moreover, such datasets are monolingual (i.e., they only include English text), preventing one from gauging the effectiveness of any given method against phishing emails in different languages. Finally, such datasets have labeling issues because oftentimes spam emails are labeled as phishing---despite being semantically very different.
    \item \textit{Lack of source code.} Only few papers release their source code (e.g.,~\cite{al2024novel}); in some cases, we found a link to empty repositories (e.g.,~\cite{mahendru2024securenet}). This is problematic because, in various cases (e.g.,~\cite{mahendru2024securenet,doshi2023comprehensive}), the dataset itself is generated by mixing different datasets (sometimes even ``enriched'' with synthetic emails~\cite{beydemir2024dynamically}, or collected from the authors' inboxes~\cite{bountakas2023helphed}. Hence, lack of open-source code not only prevents replicability of the method, but also impairs reproduction of the same testbed by future research.
    \item \textit{No clear naming.} We found that prior works may refer to a certain dataset with different names. For instance, SpamBase is referred to as ``UCI machine learning repository'' in~\cite{atawneh2023phishing} and ``UCI repository'' in~\cite{al2024novel}; whereas the experiments in~\cite{chataut2024can} are carried out on a subset of the Enron dataset---which is, however, referred to as ``the publicly available Phishing Email Detection dataset [10]``, and reference [10] (in~\cite{chataut2024can}) links to a dataset on Kaggle for which no source was specified, and only a comment from an user highlighted that the samples are taken from Enron. More generally, many papers (see~\cite{kyaw2024systematic}) refer to datasets from Kaggle (instead of the true source), preventing accurate attribution.
\end{itemize}
Due to the last points, \textit{we cannot provide a precise number} of the occurrences of any given dataset in prior work. Indeed, names are often mixed, and lack of source code prevents verifying which dataset was actually used to test a method.

\textbf{Less popular datasets.} There exist other  datasets/repositories usable to test phishing-email detectors. Among these, we mention:
Untroubled~\cite{untroubled} (used in~\cite{betts2024exploring});
PhishingPot~\cite{repository} (used in~\cite{de2024hey});
PhishBowl~\cite{phishbowl} (used in~\cite{nguyen2024utilizing});
IWSPA~\cite{IWSPA} (used in~\cite{zhao2024fewshing});
MillerSmiles~\cite{millersmiles} (used in~\cite{kulal2025phishing}); as well as the recent dataset by Chataut et al.~\cite{chataut2024enhancing}. 
Such ``less-used datasets'' may not be affected by the issues of the ``overly-used datasets''. However, we report that none of the papers we analysed that considered such datasets released their code; moreover, some datasets are ``dead'' (e.g., IWSPA has less than a dozen emails at the time of writing this article) or require a payment (e.g., only a portion of MillerSmiles is free). Finally, and importantly, some of these datasets (e.g., PhishingPot) are not designed to be ``benchmarks'', but rather are community-driven efforts to collect phishing emails, and hence they change continuously. Without source code, it is impossible to exactly replicate the same testbed. 

\begin{cooltextbox}
\textsc{\textbf{Answer to RQ1.}} Previously proposed methods are evaluated on datasets (such as SpamAssassin, SpamBase, Enron, Nazario, LingSpam) that have old (and monolingual) emails---which hardly resemble current phishing trends. Moreover, related literature often does not release their codebase: this is problematic especially given that it prevents accurate replication of the testbed.
\end{cooltextbox}



\subsection{Related Work (and Novelty)}
\label{ssec:related}

\noindent
To avoid misunderstandings and clarify our focus, let us position our paper within extant literature on phishing email detection.

Thousands of papers have tackled the problem of phishing emails. Various reviews/systematizations have analysed (or re-examined) the performance of diverse ML-based detectors, as well as the datasets used to test such detectors (see, e.g.,~\cite{das2019sok,alhuzali2025depth,salloum2022systematic,champa2024curated,kyaw2024systematic,thakur2023systematic}). We acknowledge these contributions, which is why we do not claim novelty in our aforementioned analysis (§\ref{ssec:datasets}). Indeed, our analysis serves as a scaffold to highlight that---in the phishing-email detection context---there is an ``open problem'' (i.e., the constant re-use of not-very-representative datasets) that deserves to be broadcast.

We acknowledge that our re-assessment (discussed in §\ref{sec:reassessment}), entail previously-proposed detection methods. Again, we do not claim novelty in such an evaluation. Indeed, by releasing all of our resources, our goal is to provide a solid foundation that can be used to spearhead future research in the phishing-email detection domain.

In contrast, we do claim novelty in our proposed \fmw{} framework~(§\ref{sec:framework}). Even though methods to generate synthetic dataset exists (e.g.,~\cite{mehdi2023adversarial}), we are not aware of any prior work that proposed a methodology that automatically {\small \textit{(i)}}~generates ``profiles'' of potential targets of phishing emails, and {\small \textit{(ii)}}~crafts high-quality phishing emails tailored to such profiles. We also claim originality in our \ds{} dataset (especially given that it is multilingual), whose quality has been validated with an user study (§\ref{sec:validation}). 
\section{Reassessment of Previous Detection Methods}
\label{sec:reassessment}

\noindent
In our literature analysis, we found: a shortage of publicly-available source code on phishing email detection; as well as the lack of a ``standardized'' testbed (due to the mixing of various datasets, which may have labelling issues~\cite{alhuzali2025depth,al2024novel}). Hence, to provide a foundation for future work, we carry out a comprehensive reassessment of previously proposed ML-based detection methods.

In doing so, however, we go beyond the traditional evaluation methodology of testing a model on data from the same distribution as that of the training set (done, e.g., in~\cite{alhuzali2025depth}): we will also evaluate the generalizability of the considered detectors by applying cross-evaluation methodologies~\cite{apruzzese2022cross}. We first present our setup (§\ref{ssec:setup}), then present the results (§\ref{ssec:reassessment_results}), and then draw considerations (§\ref{ssec:considerations}).

\subsection{Experimental Setup}
\label{ssec:setup}
\noindent
Recall RQ2: ``\purple{what performance do existing detectors achieve on some previously-used benchmark datasets?}'' To provide a meaningful---but humanly-feasible---answer to RQ2, we carry out our reassessment by considering eight detectors evaluated across eight datasets.

\vspace{1mm}
\textbf{Datasets (and standardization)}
We use eight publicly available email datasets, spanning across popular (e.g., Enron) and less-used ones (e.g., Chataut~\cite{chataut2024enhancing}). Table~\ref{tab:email_datasets} provides an overview of their sizes, class distributions, and sources. Importantly, to provide an evaluation that aligns with prior work, we considered ``variants'' of these datasets used by some works for which we found a dedicated repository (i.e.,~\cite{al2024novel,giri2022comparative,champa2024curated}). Note that some datasets (i.e., Enron and LingSpam) are listed twice: once for the variant by~\cite{al2024novel} and the second for~\cite{giri2022comparative}. We expect our results on these datasets to be similar (given that they are drawn from the same distribution), hence such a design choice can be used to validate our results. To enable cross-evaluations, it is necessary that all datasets are in the same format~\cite{apruzzese2022cross}. To this end, we standardize our considered datasets by clearly separating \textit{subject}, \textit{bodytext}, and \textit{label} of each email. We also cleaned the text by, e.g., removing multiple whitespaces, or HTML tags (given that not all datasets include them).

\vspace{1mm}
\textbf{Detection Methods}
We explore eight techniques pertaining to the three categories of ML-based detectors mentioned in §\ref{ssec:background}. Specifically: five feature-based classifiers---Logistic Regression (LR), Naive Bayes (NB), Random Forest (RF), Support Vector Machine (SVM), and MultiLayerPerceptron (MLP)---which rely on TF-IDF representations for the email subject and bodytext (similarly to~\cite{doshi2023comprehensive,salloum2022systematic,bountakas2021comparison}); one feature-agnostic and transformer-based model---DistilBert (DB)---which will undergo a fine-tuning process (similarly to~\cite{jamal2024improved}); and two state-of-the-art LLMs---gemini-2.0-flash and gpt-4o-mini---with zero-shot prompting (as also done in~\cite{rojas2024zero}).

\vspace{1mm}
\textbf{Evaluation Protocol}
The common practice is to test an ML model on data drawn from the same dataset used to train/fine-tune such an ML model: such a protocol prevents one from gauging the effects of combining data from different distributions. Hence, to ``maximize'' the potential of existing datasets, we explore the additional scenarios enabled by cross-evaluations~\cite{apruzzese2022cross}. Specifically, we design our evaluation around three core experiments. 
\begin{itemize}[leftmargin=*]
    \item \textit{Experiment-1}: we scrutinize whether models requiring a training/ fine-tuning phase---which typically achieve strong performance when tested on samples from the same dataset---suffer from a degraded performance when tested on samples from different datasets. This is useful to measure the generalization capabilities. 

    \item \textit{Experiment-2}: we examine an ``all-vs-one'' scenario (inspired by~\cite{apruzzese2023sok}) in which we train/fine-tune a model on data from 7 (out of 8) datasets, and we then test such a model on the left-out dataset. This is useful to see if the combination of existing datasets can cover some blind-spots, leading to models which better generalize---potentially at the expense of a lower performance on data from the same training distribution. (To our knowledge, this experiment is new in the phishing-email detection context.)

    \item \textit{Experiment-3}: we gauge how well LLMs can act as phishing-email detectors in a zero-shot context (the prompt is in our repo~\cite{repository2}).
\end{itemize}
To provide statistically significant results, we repeat our experiments five times---each time by applying a stratified 70:30 train-test split, but with a different random seed. The performance is always measured on the ``test'' partition (to avoid data leak~\cite{arp2022and}). Due to space limitations, we will report only the average F1-score (useful to combine both the true- and false-positive rate) in this paper: the complete results (e.g., accuracy, precision, recall, as well as standard deviations) are in our repository~\cite{repository2}.

\begin{table}[!htpb]
\vspace{-2mm}
\footnotesize
\centering
\begin{tabular}{l|r|r|r|c}
\hline
\textbf{Dataset Name} & \textbf{Size} & \textbf{\# Phishing} & \textbf{\# Benign} & Variant\\
\hline
CEAS~\cite{ceas} & 39126 & 21829 & 17297 & \cite{al2024novel}\\
Enron-v1~\cite{klimt2004enron} & 29569 & 13778 & 15791 & \cite{al2024novel}\\
Ling-v1~\cite{sakkis2003memory} & 2797 & 445 & 2352 & \cite{al2024novel}\\
SpamAssassin~\cite{spamassassin} & 5791 & 1704 & 4087 & \cite{al2024novel}\\
TREC~\cite{trec} & 123232 & 55291 & 67941 &\cite{champa2024curated} \\
Chataut~\cite{chataut2024enhancing} & 24583 & 19681 & 4902 & \cite{chataut2024enhancing}\\
Enron-v2~\cite{klimt2004enron} & 9601 & 4687 & 4914 & \cite{giri2022comparative} \\
Ling-v2~\cite{sakkis2003memory} & 2590 & 423 & 2167 & \cite{giri2022comparative}\\
\hline
\end{tabular}
\caption{Overview of the datasets used in our reassessment.}
\label{tab:email_datasets}
\vspace{-4mm}
\end{table}

\subsection{Results [RQ2]}\label{ssec:reassessment_results}
\noindent
Due to space limitations, we will report only the average F1-score (useful to combine both the true- and false-positive rate) in this paper: the complete results (e.g., accuracy, precision, recall, as well as standard deviations) are in our repository~\cite{repository2}

\vspace{1mm}
\textbf{Experiment-1.}
We report the results in Table~\ref{tab:cross_eval}. 
All our considered models exhibit poor cross-dataset generalization. For instance, LR and NB achieve very high F1-scores when tested on data from the same training distribution (e.g., 0.98 and 0.83 on CEAS, respectively) but their performance drops drastically in an inter-dataset context (down to 0.27 and 0.01, with average drops of 0.51 and 0.69, respectively). DistilBERT, while more robust, follows a similar trend: the performance on the same dataset is near-perfect (e.g., 1.00 on CEAS) but the performance drops on different datasets (e.g., 0.67 on SpamAssassin); yet, DistilBert has a much smaller average performance drop (e.g., fine-tuned on TREC, the average drop is 0.14---albeit the F1-score is still only 0.56 on Chataut). Such a result indicates that embedding-based models (such as DistilBERT) might have better generalization capabilities w.r.t. those that rely on vocabularies built via TF-IDF. Finally, we appreciate that all models achieve high performance when tested on data from the same ``variant'' of a given dataset (e.g., the SVM trained on Enron-v1 has 0.97 F1-score on Enron-v1, and 0.97 F1-score on Enrong v2; and viceversa): this (expected) result validates our experimental setup.

\begin{table}[!tpb]
\centering
\scriptsize
\renewcommand{\arraystretch}{1.1}
\begin{tabular}{llcccccccc|c}
\toprule
\rotatebox{90}{\textbf{Model}} & \rotatebox{90}{\textbf{Trained On}} & \rotatebox{90}{CEAS} & \rotatebox{90}{Enron-v1} & \rotatebox{90}{Ling-v1} & \rotatebox{90}{SpamAssassin} & \rotatebox{90}{TREC} & \rotatebox{90}{Chatuat} & \rotatebox{90}{Enron-v2} & \rotatebox{90}{Ling-v2} & \rotatebox{90}{\textbf{Average Drop}} \\
\midrule
\multirow{8}{*}{\rotatebox{90}{Logistic Regression}} & CEAS & \cellcolor[RGB]{146,235,143} 0.98 & \cellcolor[RGB]{185,190,137} 0.57 & \cellcolor[RGB]{212,159,132} 0.29 & \cellcolor[RGB]{209,163,133} 0.32 & \cellcolor[RGB]{174,202,138} 0.68 & \cellcolor[RGB]{185,190,137} 0.57 & \cellcolor[RGB]{184,191,137} 0.58 & \cellcolor[RGB]{213,158,132} 0.27 & 0.51 \\
 & Enron-v1 & \cellcolor[RGB]{168,209,139} 0.74 & \cellcolor[RGB]{147,234,143} 0.96 & \cellcolor[RGB]{198,175,134} 0.43 & \cellcolor[RGB]{191,183,136} 0.51 & \cellcolor[RGB]{166,212,140} 0.77 & \cellcolor[RGB]{160,219,141} 0.83 & \cellcolor[RGB]{147,234,143} 0.96 & \cellcolor[RGB]{197,176,135} 0.44 & 0.30 \\
 & Ling-v1 & \cellcolor[RGB]{196,177,135} 0.45 & \cellcolor[RGB]{180,195,137} 0.62 & \cellcolor[RGB]{152,228,142} 0.91 & \cellcolor[RGB]{170,207,139} 0.73 & \cellcolor[RGB]{188,187,136} 0.54 & \cellcolor[RGB]{206,166,133} 0.35 & \cellcolor[RGB]{180,196,137} 0.62 & \cellcolor[RGB]{151,229,142} 0.92 & 0.31 \\
 & SpamAssassin & \cellcolor[RGB]{200,173,134} 0.42 & \cellcolor[RGB]{177,199,138} 0.65 & \cellcolor[RGB]{168,209,139} 0.74 & \cellcolor[RGB]{152,227,142} 0.91 & \cellcolor[RGB]{187,188,136} 0.55 & \cellcolor[RGB]{201,171,134} 0.40 & \cellcolor[RGB]{176,200,138} 0.66 & \cellcolor[RGB]{171,206,139} 0.71 & 0.32 \\
 & TREC & \cellcolor[RGB]{159,219,141} 0.83 & \cellcolor[RGB]{151,229,142} 0.92 & \cellcolor[RGB]{174,202,138} 0.68 & \cellcolor[RGB]{170,208,139} 0.73 & \cellcolor[RGB]{149,231,143} 0.94 & \cellcolor[RGB]{183,193,137} 0.59 & \cellcolor[RGB]{151,229,142} 0.92 & \cellcolor[RGB]{177,199,138} 0.65 & 0.18 \\
 & Chatuat & \cellcolor[RGB]{171,206,139} 0.72 & \cellcolor[RGB]{179,196,138} 0.63 & \cellcolor[RGB]{216,155,131} 0.25 & \cellcolor[RGB]{196,177,135} 0.45 & \cellcolor[RGB]{180,196,137} 0.62 & \cellcolor[RGB]{144,237,143} 1.00 & \cellcolor[RGB]{177,199,138} 0.65 & \cellcolor[RGB]{214,156,132} 0.26 & 0.49 \\
 & Enron-v2 & \cellcolor[RGB]{171,206,139} 0.72 & \cellcolor[RGB]{148,232,143} 0.95 & \cellcolor[RGB]{200,172,134} 0.41 & \cellcolor[RGB]{195,179,135} 0.47 & \cellcolor[RGB]{177,199,138} 0.65 & \cellcolor[RGB]{156,223,141} 0.87 & \cellcolor[RGB]{148,232,143} 0.95 & \cellcolor[RGB]{199,174,134} 0.42 & 0.31 \\
 & Ling-v2 & \cellcolor[RGB]{192,182,135} 0.49 & \cellcolor[RGB]{177,199,138} 0.65 & \cellcolor[RGB]{150,230,142} 0.93 & \cellcolor[RGB]{170,207,139} 0.73 & \cellcolor[RGB]{186,189,136} 0.56 & \cellcolor[RGB]{202,170,134} 0.39 & \cellcolor[RGB]{176,200,138} 0.66 & \cellcolor[RGB]{150,230,142} 0.93 & 0.30 \\
\midrule
\multirow{8}{*}{\rotatebox{90}{Naive Bayes}} & CEAS & \cellcolor[RGB]{160,219,141} 0.83 & \cellcolor[RGB]{226,143,130} 0.14 & \cellcolor[RGB]{239,128,128} 0.01 & \cellcolor[RGB]{235,133,128} 0.05 & \cellcolor[RGB]{215,155,132} 0.25 & \cellcolor[RGB]{205,167,133} 0.35 & \cellcolor[RGB]{225,144,130} 0.15 & \cellcolor[RGB]{238,129,128} 0.01 & 0.69 \\
 & Enron-v1 & \cellcolor[RGB]{163,215,140} 0.79 & \cellcolor[RGB]{148,233,143} 0.95 & \cellcolor[RGB]{182,194,137} 0.60 & \cellcolor[RGB]{184,191,137} 0.58 & \cellcolor[RGB]{165,213,140} 0.78 & \cellcolor[RGB]{168,210,139} 0.75 & \cellcolor[RGB]{148,233,143} 0.96 & \cellcolor[RGB]{180,195,137} 0.62 & 0.23 \\
 & Ling-v1 & \cellcolor[RGB]{173,204,139} 0.70 & \cellcolor[RGB]{171,206,139} 0.71 & \cellcolor[RGB]{146,234,143} 0.97 & \cellcolor[RGB]{188,187,136} 0.54 & \cellcolor[RGB]{175,202,138} 0.67 & \cellcolor[RGB]{171,206,139} 0.71 & \cellcolor[RGB]{170,207,139} 0.72 & \cellcolor[RGB]{147,233,143} 0.96 & 0.25 \\
 & SpamAssassin & \cellcolor[RGB]{175,201,138} 0.67 & \cellcolor[RGB]{173,204,139} 0.70 & \cellcolor[RGB]{180,196,137} 0.62 & \cellcolor[RGB]{148,232,143} 0.95 & \cellcolor[RGB]{174,202,138} 0.68 & \cellcolor[RGB]{196,177,135} 0.45 & \cellcolor[RGB]{172,205,139} 0.71 & \cellcolor[RGB]{179,197,138} 0.63 & 0.31 \\
 & TREC & \cellcolor[RGB]{164,214,140} 0.79 & \cellcolor[RGB]{153,227,142} 0.90 & \cellcolor[RGB]{160,219,141} 0.83 & \cellcolor[RGB]{163,215,140} 0.79 & \cellcolor[RGB]{155,224,142} 0.88 & \cellcolor[RGB]{190,184,136} 0.51 & \cellcolor[RGB]{153,227,142} 0.90 & \cellcolor[RGB]{161,218,141} 0.82 & 0.08 \\
 & Chatuat & \cellcolor[RGB]{174,203,138} 0.69 & \cellcolor[RGB]{179,196,138} 0.62 & \cellcolor[RGB]{213,158,132} 0.27 & \cellcolor[RGB]{196,177,135} 0.45 & \cellcolor[RGB]{182,194,137} 0.60 & \cellcolor[RGB]{145,236,143} 0.98 & \cellcolor[RGB]{178,198,138} 0.64 & \cellcolor[RGB]{213,158,132} 0.28 & 0.48 \\
 & Enron-v2 & \cellcolor[RGB]{163,215,140} 0.79 & \cellcolor[RGB]{148,233,143} 0.96 & \cellcolor[RGB]{182,193,137} 0.59 & \cellcolor[RGB]{184,191,137} 0.58 & \cellcolor[RGB]{165,213,140} 0.78 & \cellcolor[RGB]{167,210,140} 0.75 & \cellcolor[RGB]{148,233,143} 0.96 & \cellcolor[RGB]{180,195,137} 0.62 & 0.23 \\
 & Ling-v2 & \cellcolor[RGB]{173,204,139} 0.70 & \cellcolor[RGB]{171,206,139} 0.71 & \cellcolor[RGB]{146,235,143} 0.97 & \cellcolor[RGB]{187,187,136} 0.54 & \cellcolor[RGB]{175,202,138} 0.68 & \cellcolor[RGB]{171,206,139} 0.71 & \cellcolor[RGB]{170,208,139} 0.73 & \cellcolor[RGB]{148,233,143} 0.96 & 0.24 \\
\midrule
\multirow{8}{*}{\rotatebox{90}{Random Forest}} & CEAS & \cellcolor[RGB]{144,237,143} 0.99 & \cellcolor[RGB]{190,185,136} 0.52 & \cellcolor[RGB]{217,153,131} 0.23 & \cellcolor[RGB]{228,140,129} 0.12 & \cellcolor[RGB]{184,192,137} 0.58 & \cellcolor[RGB]{191,183,136} 0.51 & \cellcolor[RGB]{188,186,136} 0.53 & \cellcolor[RGB]{219,151,131} 0.21 & 0.60 \\
 & Enron-v1 & \cellcolor[RGB]{169,208,139} 0.73 & \cellcolor[RGB]{146,235,143} 0.98 & \cellcolor[RGB]{197,176,135} 0.44 & \cellcolor[RGB]{188,187,136} 0.54 & \cellcolor[RGB]{163,216,140} 0.80 & \cellcolor[RGB]{165,213,140} 0.78 & \cellcolor[RGB]{145,236,143} 0.99 & \cellcolor[RGB]{196,178,135} 0.46 & 0.30 \\
 & Ling-v1 & \cellcolor[RGB]{195,179,135} 0.47 & \cellcolor[RGB]{171,206,139} 0.72 & \cellcolor[RGB]{146,235,143} 0.98 & \cellcolor[RGB]{171,206,139} 0.71 & \cellcolor[RGB]{183,192,137} 0.59 & \cellcolor[RGB]{199,174,134} 0.42 & \cellcolor[RGB]{171,207,139} 0.72 & \cellcolor[RGB]{145,236,143} 0.99 & 0.32 \\
 & SpamAssassin & \cellcolor[RGB]{195,178,135} 0.46 & \cellcolor[RGB]{173,203,139} 0.69 & \cellcolor[RGB]{174,202,138} 0.68 & \cellcolor[RGB]{147,234,143} 0.97 & \cellcolor[RGB]{179,197,138} 0.63 & \cellcolor[RGB]{201,171,134} 0.40 & \cellcolor[RGB]{172,204,139} 0.70 & \cellcolor[RGB]{175,201,138} 0.67 & 0.36 \\
 & TREC & \cellcolor[RGB]{159,220,141} 0.84 & \cellcolor[RGB]{150,231,142} 0.94 & \cellcolor[RGB]{184,191,137} 0.58 & \cellcolor[RGB]{165,213,140} 0.77 & \cellcolor[RGB]{146,234,143} 0.97 & \cellcolor[RGB]{188,187,136} 0.54 & \cellcolor[RGB]{150,230,142} 0.94 & \cellcolor[RGB]{185,190,137} 0.57 & 0.23 \\
 & Chatuat & \cellcolor[RGB]{171,206,139} 0.72 & \cellcolor[RGB]{178,198,138} 0.64 & \cellcolor[RGB]{213,158,132} 0.28 & \cellcolor[RGB]{196,178,135} 0.45 & \cellcolor[RGB]{180,196,137} 0.62 & \cellcolor[RGB]{144,237,143} 1.00 & \cellcolor[RGB]{176,200,138} 0.66 & \cellcolor[RGB]{212,158,132} 0.28 & 0.48 \\
 & Enron-v2 & \cellcolor[RGB]{171,206,139} 0.71 & \cellcolor[RGB]{146,235,143} 0.97 & \cellcolor[RGB]{200,173,134} 0.41 & \cellcolor[RGB]{189,185,136} 0.52 & \cellcolor[RGB]{164,214,140} 0.79 & \cellcolor[RGB]{163,215,140} 0.80 & \cellcolor[RGB]{146,234,143} 0.97 & \cellcolor[RGB]{199,174,134} 0.43 & 0.31 \\
 & Ling-v2 & \cellcolor[RGB]{192,182,135} 0.50 & \cellcolor[RGB]{170,207,139} 0.72 & \cellcolor[RGB]{145,236,143} 0.99 & \cellcolor[RGB]{171,206,139} 0.71 & \cellcolor[RGB]{182,193,137} 0.60 & \cellcolor[RGB]{197,176,135} 0.44 & \cellcolor[RGB]{170,207,139} 0.73 & \cellcolor[RGB]{145,235,143} 0.98 & 0.31 \\
\midrule
\multirow{8}{*}{\rotatebox{90}{Support Vector Machine}} & CEAS & \cellcolor[RGB]{145,236,143} 0.99 & \cellcolor[RGB]{181,194,137} 0.61 & \cellcolor[RGB]{212,159,132} 0.28 & \cellcolor[RGB]{211,161,132} 0.30 & \cellcolor[RGB]{171,206,139} 0.72 & \cellcolor[RGB]{186,188,136} 0.55 & \cellcolor[RGB]{180,196,137} 0.62 & \cellcolor[RGB]{212,159,132} 0.28 & 0.50 \\
 & Enron-v1 & \cellcolor[RGB]{166,212,140} 0.77 & \cellcolor[RGB]{146,235,143} 0.97 & \cellcolor[RGB]{197,176,135} 0.44 & \cellcolor[RGB]{190,184,136} 0.51 & \cellcolor[RGB]{165,213,140} 0.78 & \cellcolor[RGB]{160,219,141} 0.83 & \cellcolor[RGB]{146,235,143} 0.97 & \cellcolor[RGB]{196,177,135} 0.45 & 0.29 \\
 & Ling-v1 & \cellcolor[RGB]{199,173,134} 0.42 & \cellcolor[RGB]{173,204,139} 0.70 & \cellcolor[RGB]{147,234,143} 0.96 & \cellcolor[RGB]{165,213,140} 0.77 & \cellcolor[RGB]{182,194,137} 0.60 & \cellcolor[RGB]{203,170,134} 0.38 & \cellcolor[RGB]{172,205,139} 0.70 & \cellcolor[RGB]{149,231,143} 0.94 & 0.32 \\
 & SpamAssassin & \cellcolor[RGB]{194,180,135} 0.47 & \cellcolor[RGB]{174,202,138} 0.68 & \cellcolor[RGB]{170,207,139} 0.72 & \cellcolor[RGB]{149,232,143} 0.94 & \cellcolor[RGB]{183,193,137} 0.59 & \cellcolor[RGB]{200,173,134} 0.41 & \cellcolor[RGB]{173,203,139} 0.69 & \cellcolor[RGB]{174,203,138} 0.68 & 0.34 \\
 & TREC & \cellcolor[RGB]{159,220,141} 0.84 & \cellcolor[RGB]{149,231,143} 0.94 & \cellcolor[RGB]{172,205,139} 0.70 & \cellcolor[RGB]{168,209,139} 0.74 & \cellcolor[RGB]{148,232,143} 0.95 & \cellcolor[RGB]{182,193,137} 0.60 & \cellcolor[RGB]{150,230,142} 0.94 & \cellcolor[RGB]{175,201,138} 0.67 & 0.18 \\
 & Chatuat & \cellcolor[RGB]{171,206,139} 0.72 & \cellcolor[RGB]{178,198,138} 0.64 & \cellcolor[RGB]{213,158,132} 0.27 & \cellcolor[RGB]{196,177,135} 0.45 & \cellcolor[RGB]{180,196,137} 0.62 & \cellcolor[RGB]{144,237,143} 1.00 & \cellcolor[RGB]{176,200,138} 0.66 & \cellcolor[RGB]{213,158,132} 0.28 & 0.48 \\
 & Enron-v2 & \cellcolor[RGB]{170,208,139} 0.73 & \cellcolor[RGB]{147,234,143} 0.97 & \cellcolor[RGB]{201,172,134} 0.40 & \cellcolor[RGB]{194,180,135} 0.48 & \cellcolor[RGB]{171,206,139} 0.72 & \cellcolor[RGB]{156,223,141} 0.87 & \cellcolor[RGB]{147,233,143} 0.96 & \cellcolor[RGB]{200,173,134} 0.41 & 0.31 \\
 & Ling-v2 & \cellcolor[RGB]{197,176,135} 0.44 & \cellcolor[RGB]{172,205,139} 0.70 & \cellcolor[RGB]{146,235,143} 0.98 & \cellcolor[RGB]{166,212,140} 0.77 & \cellcolor[RGB]{180,195,137} 0.62 & \cellcolor[RGB]{201,172,134} 0.40 & \cellcolor[RGB]{171,206,139} 0.71 & \cellcolor[RGB]{147,234,143} 0.97 & 0.31 \\
\midrule
\multirow{8}{*}{\rotatebox{90}{Multi-Layer Perceptron}} & CEAS & \cellcolor[RGB]{144,237,143} 0.99 & \cellcolor[RGB]{173,203,139} 0.69 & \cellcolor[RGB]{201,171,134} 0.40 & \cellcolor[RGB]{195,178,135} 0.46 & \cellcolor[RGB]{165,213,140} 0.77 & \cellcolor[RGB]{188,187,136} 0.54 & \cellcolor[RGB]{173,204,139} 0.69 & \cellcolor[RGB]{200,172,134} 0.41 & 0.43 \\
 & Enron-v1 & \cellcolor[RGB]{166,212,140} 0.76 & \cellcolor[RGB]{146,235,143} 0.98 & \cellcolor[RGB]{182,194,137} 0.60 & \cellcolor[RGB]{182,193,137} 0.60 & \cellcolor[RGB]{159,219,141} 0.83 & \cellcolor[RGB]{169,208,139} 0.74 & \cellcolor[RGB]{145,236,143} 0.98 & \cellcolor[RGB]{181,194,137} 0.61 & 0.25 \\
 & Ling-v1 & \cellcolor[RGB]{187,187,136} 0.54 & \cellcolor[RGB]{176,200,138} 0.66 & \cellcolor[RGB]{146,234,143} 0.97 & \cellcolor[RGB]{163,215,140} 0.79 & \cellcolor[RGB]{181,195,137} 0.61 & \cellcolor[RGB]{200,173,134} 0.42 & \cellcolor[RGB]{175,202,138} 0.67 & \cellcolor[RGB]{147,234,143} 0.97 & 0.30 \\
 & SpamAssassin & \cellcolor[RGB]{178,198,138} 0.64 & \cellcolor[RGB]{172,205,139} 0.70 & \cellcolor[RGB]{168,210,139} 0.75 & \cellcolor[RGB]{146,234,143} 0.97 & \cellcolor[RGB]{174,202,138} 0.68 & \cellcolor[RGB]{199,174,134} 0.43 & \cellcolor[RGB]{171,206,139} 0.71 & \cellcolor[RGB]{169,208,139} 0.73 & 0.31 \\
 & TREC & \cellcolor[RGB]{161,218,141} 0.82 & \cellcolor[RGB]{148,232,143} 0.95 & \cellcolor[RGB]{176,201,138} 0.66 & \cellcolor[RGB]{169,208,139} 0.74 & \cellcolor[RGB]{146,235,143} 0.97 & \cellcolor[RGB]{185,190,137} 0.57 & \cellcolor[RGB]{148,232,143} 0.95 & \cellcolor[RGB]{177,199,138} 0.65 & 0.21 \\
 & Chatuat & \cellcolor[RGB]{171,206,139} 0.72 & \cellcolor[RGB]{178,198,138} 0.64 & \cellcolor[RGB]{213,158,132} 0.28 & \cellcolor[RGB]{196,178,135} 0.45 & \cellcolor[RGB]{180,196,137} 0.62 & \cellcolor[RGB]{144,237,143} 1.00 & \cellcolor[RGB]{176,200,138} 0.66 & \cellcolor[RGB]{213,158,132} 0.28 & 0.48 \\
 & Enron-v2 & \cellcolor[RGB]{170,208,139} 0.73 & \cellcolor[RGB]{146,235,143} 0.98 & \cellcolor[RGB]{180,196,137} 0.62 & \cellcolor[RGB]{199,174,134} 0.42 & \cellcolor[RGB]{177,199,138} 0.65 & \cellcolor[RGB]{167,210,140} 0.75 & \cellcolor[RGB]{146,234,143} 0.97 & \cellcolor[RGB]{178,198,138} 0.64 & 0.29 \\
 & Ling-v2 & \cellcolor[RGB]{189,186,136} 0.53 & \cellcolor[RGB]{176,200,138} 0.66 & \cellcolor[RGB]{145,236,143} 0.98 & \cellcolor[RGB]{164,214,140} 0.79 & \cellcolor[RGB]{181,194,137} 0.61 & \cellcolor[RGB]{200,173,134} 0.41 & \cellcolor[RGB]{175,201,138} 0.67 & \cellcolor[RGB]{147,233,143} 0.96 & 0.30 \\
\midrule
\multirow{8}{*}{\rotatebox{90}{DistilBERT}} & CEAS & \cellcolor[RGB]{144,237,143} 1.00 & \cellcolor[RGB]{159,219,141} 0.83 & \cellcolor[RGB]{180,196,137} 0.62 & \cellcolor[RGB]{175,201,138} 0.67 & \cellcolor[RGB]{160,219,141} 0.83 & \cellcolor[RGB]{186,189,136} 0.56 & \cellcolor[RGB]{159,220,141} 0.84 & \cellcolor[RGB]{185,190,137} 0.57 & 0.30 \\
 & Enron-v1 & \cellcolor[RGB]{163,216,140} 0.80 & \cellcolor[RGB]{144,237,143} 0.99 & \cellcolor[RGB]{184,191,137} 0.58 & \cellcolor[RGB]{182,193,137} 0.60 & \cellcolor[RGB]{155,224,142} 0.88 & \cellcolor[RGB]{166,212,140} 0.77 & \cellcolor[RGB]{144,237,143} 1.00 & \cellcolor[RGB]{187,188,136} 0.55 & 0.26 \\
 & Ling-v1 & \cellcolor[RGB]{162,217,140} 0.81 & \cellcolor[RGB]{171,206,139} 0.71 & \cellcolor[RGB]{144,237,143} 1.00 & \cellcolor[RGB]{189,185,136} 0.52 & \cellcolor[RGB]{173,204,139} 0.69 & \cellcolor[RGB]{168,210,139} 0.75 & \cellcolor[RGB]{170,207,139} 0.72 & \cellcolor[RGB]{144,237,143} 1.00 & 0.25 \\
 & SpamAssassin & \cellcolor[RGB]{159,220,141} 0.84 & \cellcolor[RGB]{162,216,140} 0.81 & \cellcolor[RGB]{168,209,139} 0.74 & \cellcolor[RGB]{145,236,143} 0.98 & \cellcolor[RGB]{162,216,140} 0.80 & \cellcolor[RGB]{186,189,136} 0.56 & \cellcolor[RGB]{161,217,141} 0.81 & \cellcolor[RGB]{166,212,140} 0.77 & 0.22 \\
 & TREC & \cellcolor[RGB]{156,223,141} 0.86 & \cellcolor[RGB]{145,236,143} 0.98 & \cellcolor[RGB]{154,225,142} 0.89 & \cellcolor[RGB]{159,220,141} 0.84 & \cellcolor[RGB]{144,237,143} 0.99 & \cellcolor[RGB]{186,189,136} 0.56 & \cellcolor[RGB]{145,236,143} 0.98 & \cellcolor[RGB]{156,223,141} 0.87 & 0.14 \\
 & Chatuat & \cellcolor[RGB]{171,206,139} 0.71 & \cellcolor[RGB]{178,198,138} 0.64 & \cellcolor[RGB]{213,158,132} 0.28 & \cellcolor[RGB]{196,177,135} 0.45 & \cellcolor[RGB]{180,196,137} 0.62 & \cellcolor[RGB]{144,238,144} 1.00 & \cellcolor[RGB]{176,200,138} 0.66 & \cellcolor[RGB]{213,158,132} 0.28 & 0.48 \\
 & Enron-v2 & \cellcolor[RGB]{160,219,141} 0.83 & \cellcolor[RGB]{144,237,143} 0.99 & \cellcolor[RGB]{189,185,136} 0.52 & \cellcolor[RGB]{186,189,136} 0.56 & \cellcolor[RGB]{159,220,141} 0.84 & \cellcolor[RGB]{162,216,140} 0.80 & \cellcolor[RGB]{144,237,143} 0.99 & \cellcolor[RGB]{192,182,135} 0.49 & 0.27 \\
 & Ling-v2 & \cellcolor[RGB]{162,217,140} 0.81 & \cellcolor[RGB]{173,204,139} 0.69 & \cellcolor[RGB]{144,237,143} 1.00 & \cellcolor[RGB]{190,185,136} 0.52 & \cellcolor[RGB]{174,202,138} 0.68 & \cellcolor[RGB]{169,209,139} 0.74 & \cellcolor[RGB]{173,204,139} 0.69 & \cellcolor[RGB]{145,236,143} 0.98 & 0.25 \\
\bottomrule
\end{tabular}
\caption{Cross-evaluation results (Experiment-1). \textmd{We report the (averaged over 5 trials) F1-scores achieved by each model when trained on each dataset (rows) and tested on any dataset (columns). The last column shows the average drop in F1-score when testing on other datasets compared to the diagonal.}}
\label{tab:cross_eval}
\vspace{-8mm}
\end{table}

\begin{figure}[t]
\vspace{-4mm}
    \centering
    \includegraphics[width=0.8\linewidth]{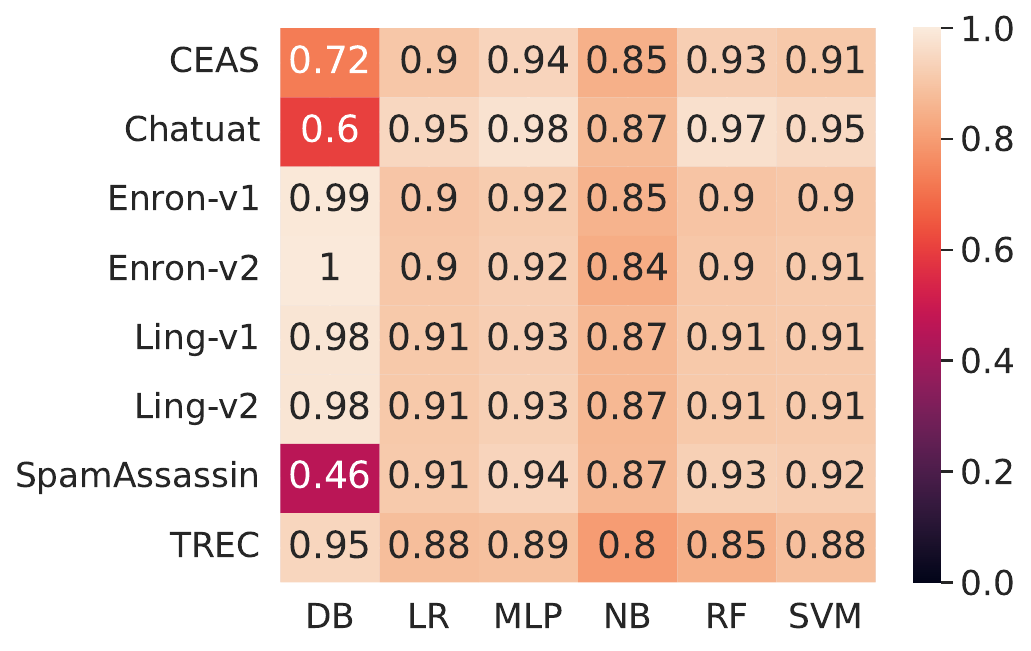}
    \vspace{-4mm}
    \caption{All-vs-one results (Experiment-2). \textmd{We train each model (columns) on seven datasets, and test it on the left-out dataset (rows).}}
    \label{fig:allvsone}
\end{figure}

\vspace{1mm}
\textbf{Experiment-2.}
We report the results of the models trained on 7 datasets and tested on the left-out dataset in Fig.~\ref{fig:allvsone}. Perhaps surprisingly, the MLP (which uses TF-IDF) has the most consistent performance (from 0.89 on TREC, to 0.98 on Chataut) across all datasets: in comparison, DistilBERT has much wider margins (from 0.46 on SpamAssassin, to 1.00 on Enron-v2). This indicates that, if there is availability of large amounts of data from different distributions, it may be wiser to use ML models reliant on TF-IDF (such as the simple MLP or RF) rather than embedding-based models (such as DistilBert). Finally, to gauge if an expanded training set leads to any performance degradation on data from the same distribution, we measure the (averaged) F1-score of the models trained on each of these all-but-one combinations of datasets and compare it with the (averaged) F1-score of the ``diagonals'' in Table~\ref{tab:cross_eval}. The results are in Table~\ref{tab:internal}. We can see a substantial drop in the same-dataset performance---which is an expected result. Indeed, this is the cost of developing ML models with better generalizability.

\begin{table}[!htpb]
    \footnotesize
    \centering
    \begin{tabular}{c|c|c} \toprule
         \textbf{\textit{Model}} & \textit{\textbf{Single}}&  \textbf{\textit{AllvsOne}}\\ \midrule
         LR & 0.95 & 0.90\\
         NB & 0.94 & 0.85\\
         MLP & 0.98 & 0.92\\
         RF & 0.98 & 0.91\\
         SVM & 0.97 & 0.91\\ 
         DB & 0.99 & 0.95\\\bottomrule
    \end{tabular}
    \caption{Performance on data from the same training set. \textmd{We report the average F1-score achieved by the models in Experiment-2 (AllvsOne column) and those in Experiment-1 (Single column) when tested on data from the same training set.}}
    \label{tab:internal}
    \vspace{-8mm}
\end{table}

\begin{figure}[t]
    \centering
    \includegraphics[width=.6\linewidth]{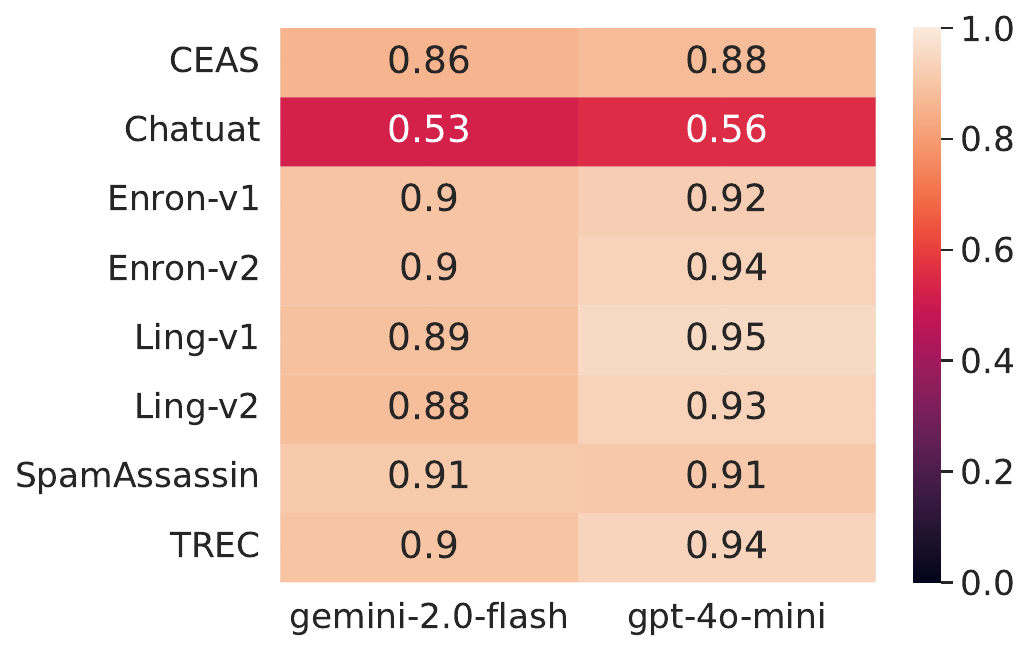}
    \vspace{-3mm}
    \caption{LLM performance (Experiment-3).}
    \label{fig:llm}
    \vspace{-3mm}
\end{figure}

\paragraph{\textbf{Experiment-3}}
We report the results of the LLMs in Figure~\ref{fig:llm}.
Both Gemini-2.0-Flash and gpt-4o-mini achieve F1-scores above 0.86, with the only exception being on the Chataut dataset. Intriguingly, even DistilBERT (which also uses transformers) struggled in the all-vs-one scenario on the Chataut dataset (see Fig.~\ref{fig:allvsone}). This result may indicate that the Chataut dataset is either substantially different from the others, or that there may be some labelling issues (e.g., benign emails labeled as phishing, or vice-versa).

\begin{cooltextbox}
\textsc{\textbf{Answer to RQ2.}} We confirm that models requiring a training phase achieve near-perfect performance when tested on data from the same dataset. Yet, such models struggle when tested on data from other datasets---but DistilBERT seems to have a better generalization power. Zero-shot-prompted LLMs exhibit high F1-scores when tasked to detect phishing emails from our considered datasets, but the performance on Chataut is poor.
\end{cooltextbox}


\subsection{Considerations}
\label{ssec:considerations}

\noindent
Let us discuss and position our findings within extant work.

First, our reassessment includes five repeated trials for each experiment, and each trial entails having all methods tested on the same test portion of each dataset. Such a 
procedure therefore enables one to carry out \textit{statistical analyses} to gauge if any given method is better than another. For instance, in the same-dataset setting (i.e., the diagonal in Table~\ref{tab:cross_eval}) one would find that the two models with highest average F1-score across all datasets are the MLP 0.977 (avg: 0.977, std: 0.016) and the RF (avg: 0.978, std: 0.011); these results are produced by 40 evaluations (5 trials on 8 datasets): a t-test reveals that these two methods can be considered as statistically equivalent (\smamath{p}\smamath{<}\smamath{.05}). Notably, such statistical tests are not mentioned even in popular reviews (e.g.,~\cite{das2019sok,alhuzali2025depth,salloum2022systematic,champa2024curated,kyaw2024systematic,thakur2023systematic}). Our repository~\cite{repository2} includes all the data used to compute such tests.

Second, we showed that cross-evaluations are a good way to test the generalizability of prior methods, since models with high performance on the same dataset exhibit a substantial drop when tested on a different dataset. Yet, there is a problem: mixing different datasets may create \textit{temporal bias} (see~\cite{arp2022and}) because the samples in the training set may come from datasets collected after those used in the test set (e.g., Chataut was released in 2024, whereas SpamAssassin in 2005). Moreover, existing datasets do not reflect current phishing trends (e.g., they do not have LLM-generated emails, which are popular today~\cite{proofpoint2024phish, weinz2025impact}) and are mostly monolingual.

The aforementioned problems led us to RQ3 (``\textit{\purple{what is a way to overcome the shortcomings of existing datasets---without raising privacy concerns?}}''), the answer of which is in the next section.

\section{Proposed \fmw[\large]~Framework}\label{sec:framework}

\noindent
We now present our proposed \fmw{} (short for ``Email Phishing Generator'') framework---the answer to RQ3. In principle, we want to provide a tool that enables researchers to {\small \textit{(i)}}~automatically generate a large corpus of emails to develop/test phishing-email detectors, which {\small \textit{(ii)}}~are of high-quality and can reflect current trends, such as being LLM-written, while {\small \textit{(iii)}}~not raising privacy concerns related to using emails from personal inboxes. 

We first describe our \fmw{} framework (§\ref{ssec:overview} to §\ref{ssec:module2}), and then use it to generate our proposed \ds{} dataset (§\ref{ssec:ds}).

\subsection{Overview}
\label{ssec:overview}
\noindent
\fmw{} is designed as a modular pipeline for generating realistic, diverse, and context-aware synthetic emails---including both benign (ham) and malicious (phishing) examples. It leverages LLMs to simulate plausible organizational contexts and communications, enabling the creation of high-quality datasets for phishing detection research. The framework is composed of two primary modules:
\begin{enumerate}[leftmargin=*]
    \item \textit{Profile Generation.} This module uses an LLM to generate synthetic company and user profiles from minimal input prompts or high-level descriptors (e.g., industry sector, company size, or regional context). Given these inputs, the LLM constructs realistic corporate structures and employee personas, including names, roles, departments, and hobbies. These profiles serve as the foundation for creating personalized and coherent email communications in subsequent steps.
    \item \textit{Email Generation.} Building on the profiles produced in the previous module, this module employs the LLM to craft emails that reflect realistic communications. It produces both legitimate (ham) messages—such as meeting requests, announcements, or transactional updates—and phishing emails tailored to the recipient’s role and organizational context. Phishing variants cover a range of attack vectors, including credential harvesting, malicious attachments, and social-engineering scams.
\end{enumerate}
The combination of a role-aware context and varied phishing strategies allows the generation of challenging and diverse training data.
\begin{figure*}[!htpb]
    \centering
    \includegraphics[width=0.7\linewidth]{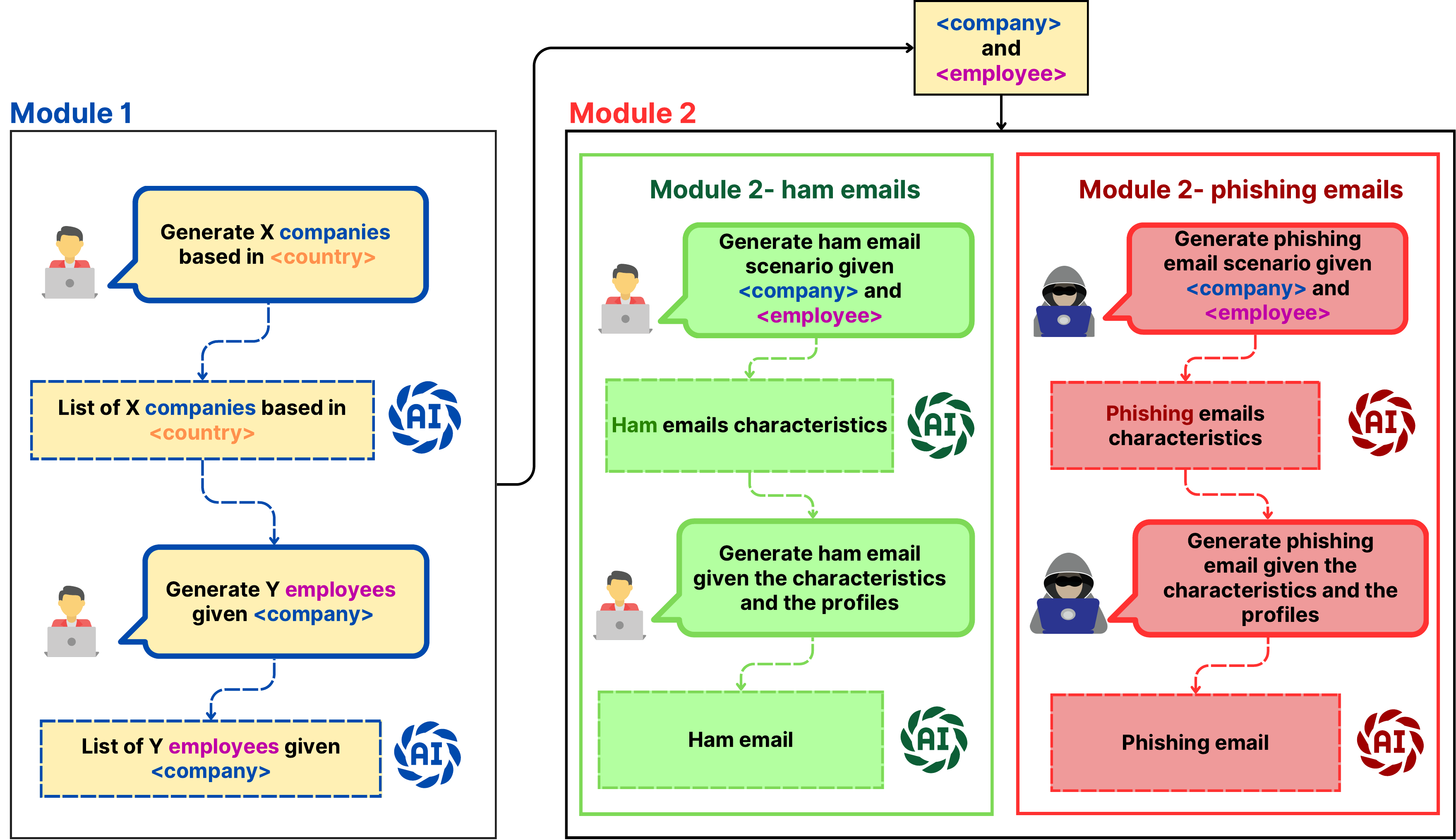}
    \vspace{-3mm}
    \caption{\textbf{Overview of \fmw{}.}. \textmd{The framework is composed of two modules: Module 1 generates synthetic company and employee profiles based on high-level input prompts; Module 2 takes these profiles as input and uses an LLM to generate realistic emails, including both benign and phishing variants, tailored to the organizational context and user roles.}}
    \label{fig:enter-label}
    \vspace{-3mm}
\end{figure*}


\subsection{Module 1: Profile Generation}
\noindent
This module generates realistic organizational contexts based on a user-defined country. The process unfolds in two steps:
\begin{enumerate}[leftmargin=*]
    \item \textit{Company Generation}. Given a selected $Country$ (e.g. United States, Italy), the framework generates $X$ synthetic companies that reflect the local economic and cultural context. This step influences the language and tone of emails produced in later stages. Companies are described using meaningful attributes such as sector, size, and region. Prompt~\ref{prompt.company-gen} shows a simplified version of the prompt for generating the companies based on a country defined by the user
    \item \textit{Employee Generation}. For each company, the system creates $Y$ synthetic employee profiles that emulate realistic corporate diversity. Each profile includes key attributes, such as role, job title, seniority, and active projects, that condition the email generation process. The Prompt~\ref{prompt.profile-gen} shows a simplified version of the prompt to generate the user profile. 
\end{enumerate}
Note that $country$, $X$, $Y$ are parameters of our framework.

\begin{prompt}[sharp corners, boxrule=0.8mm, label = {prompt.company-gen}]{Company Generation (\textit{Country, X})}
Generate $X$ fictional companies based in $Country$, reflecting a realistic distribution in regional diversity, income, and sector. Each company should have these characteristics: Name, founding year, product / services, background history, headquarter location, number of employees, annual revenue, main clients, extent of the business. 
\end{prompt}

\begin{prompt}[sharp corners, boxrule=0.8mm, label = {prompt.profile-gen}]{Profile Generation (\textit{CompanyProfile, Y})}
Generate $Y$ realistic employee profiles for a company, specifically create profiles of those who are likely to use a personal computer in their daily work. Ensure that a mixed but realistic set of employees is generated that covers senior, mid-level, and junior roles. \\
Company profile: \{$CompanyProfile$\}\\
Each user profile must include: Name, Gender, Age, Birthplace, Education, Languages, Role, Current Projects, Time employed in the company, Tech Proficiency, 
 Hobbies, Social Media. 
\end{prompt}

\subsection{Module 2: Email generation}
\label{ssec:module2}

\noindent
This module generates realistic emails—both benign (ham) and phishing—based on the synthetic company and employee profiles from Module 1. The generation process includes two main steps:
\begin{itemize}[leftmargin=*]
    \item \textit{Email Scenario Generation}. Using the company and employee profiles as input, the framework produces $N$ email scenarios, capturing key contextual attributes (e.g., topic, urgency, tone) that shape the content and intent of the email. 
    \item \textit{Email Content Generation}. Given the company, employee, and associated scenario, the LLM generates $N$ full email texts tailored to each context.
\end{itemize}
Therefore, as $X$ is a parameter, the framework will produce $X \times Y \times N$ emails for each country. In this module, the email scenario and content is highly dependent on the focus on the email, and we designed one for each of the ham and phishing categories scenarios.
For instance, as shown in Prompt~\ref{prompt.scenario}, one of the setting can be $type$ that can be a legitimate or phishing email. Furthermore, the $EmailTraits$ defines the type of characteristics we need to generate the email. For the legitimate scenario: content description, sender, tone, style, length, and receiver info.   
For the phishing scenario: phishing type, customization level, objective, impersonated identity, method, social engineering technique, tone and style, length.

\begin{prompt}[sharp corners, boxrule=0.8mm, label = {prompt.scenario}]{Scenario Generation (\textit{type, N, CompanyProfile, UserProfile, EmailTraits})}
Generate the following characteristics for $N$ $type$ emails that the employee might receive based on $CompanyProfile$ and $UserProfiles$. Each email profile must contain: $EmailTraits$. 
\end{prompt}
\addtocounter{promptcounter}{1}

\subsection{Proposed~\ds[\normalsize] Dataset}
\label{ssec:ds}
\noindent
We now demonstrate the application of our \fmw{} framework, and use it to generate our proposed \ds{} dataset. We set ourself the objective to generate emails in three languages: English, Italian, and German. 

For each language, we generate a series of company profiles for the relevant country (for English, we use both the UK and the US). Then, for each company, we generated 5 profiles of likely employees. Finally, for each employee, we generated 10 emails, five legitimate and 5 phishing. 
In total, we generated the following companies, spanning from the various countries: 250 between the UK and the US, 60 in Italy, and 50 in Germany. Note that as \fmw{} involves the utilization of LLM, some email generation procedures were erroneous and have been discarded (e.g., we expected a JSON as an output, and the LLM answer was not properly formatted). We utilized GPT-4o-mini from OpenAI as an LLM.

The \ds{} dataset consists of a total of 16,616 emails. The class distribution is balanced, with emails equally distributed between legitimate and phishing. In terms of language, the majority of emails are in English (69.22\%), followed by Italian (16.26\%) and German (14.14\%). This multilingual composition highlights the utility of \ds{} for language-agnostic phishing detection strategies.

We now show an output example for an Italian company.
\vspace{-1mm}
\begin{lstlisting}
{"company_name": "Fabbri Tech Automazione", "establishment_year": "2003", "offered_products_services": "Industrial automation systems and robotics", "company_details": "Fabbri Tech reflects the growing trend towards automation in Italy's manufacturing sector, providing cutting-edge robotics solutions. Known for custom-tailored software and engineering services, it serves a variety of industries, establishing a strong reputation in the European market.", "headquarters_location": "Modena, Emilia-Romagna", "number_of_employees": "320", "annual_revenue": "EUR60 million", "main_consumer": "Manufacturers across various sectors including automotive and food processing", "affairs_extent": "International, with clients primarily in Europe and Asia."}
\end{lstlisting}
\vspace{-1mm}
The model generates a `Tech Automatation' company based in `Emilia-Romagna' (a well-known region famous for this type of company), with an annual revenue of "€60 million". The LLM uses this type of information to generate a likely employee. 
\vspace{-1mm}
\begin{lstlisting}
{"name": "Marco Bianchi", "gender": "M", "age": 29, "birthplace": "Florence, Italy", "qualifications": "Bachelor's in Electromechanical Engineering, Universita degli Studi di Firenze", "languages": "Italian (Native), English (Intermediate), Spanish (Basic)", "job_title": "Junior Project Coordinator", "current_project": "Assisting in the coordination and planning of projects involving automated systems in automotive manufacturing", "time_employed": "2 years", "tech_proficiency": "Intermediate", "hobbies": "Traveling, Video gaming", "social_media": "LinkedIn, Facebook"}
\end{lstlisting}
\vspace{-1mm}
The employee, Marco Bianchi, has a bachelor's in Electromechanical Engineering, and his profile fits what is described by the company. Finally, with both information, we see two example of emails generated, one benign and one malicious. 

\vspace{2mm}

\begin{emailBox}{Scheduling a Call for Supply Chain Adjustments }\label{email:benign}
{\scriptsize
Dear Marco, I hope this message finds you well. We need to schedule a video call to discuss some adjustments and potential delays in the supply chain affecting our current project components. Could you please inform me of your availability this week? Looking forward to hearing from you. Best regards, TomJohnson 
}
\end{emailBox}
\vspace{-1mm}
{\noindent\small\textbf{Email~\ref{email:benign}.} Illustrative example of a benign email in \ds{}.}

\vspace{2mm}

\begin{emailBox}{Urgente: Verifica delle Credenziali dell’Account}\label{email:malicious}
{
{\scriptsize
Ciao Marco, Ti scrivo per conto del tuo manager per richiedere un’urgenteverifica delle tue credenziali aziendali. È molto importante che tuproceda al controllo immediato della correttezza delle informazioni d’accesso personali. Si prega di seguire il link di verifica di seguito e aggiornare qualsiasi informazione necessaria quanto prima:«link» Grazieperla tua collaborazione. Cordiali saluti, Federica Rossi Responsabile IT Fabbri Tech Automazione}}
\end{emailBox}
\vspace{-1mm}
{\noindent\small\textbf{Email~\ref{email:malicious}.} Illustrative example of a malicious email in \ds{}.}

\vspace{2mm}

By analyzing Marcos' email, we noticed that as the company operates in international markets, incoming emails are written in both Italian and English. We believe that this automation provided by LLM, provides a realistic scenario of working experience.

\section{Benchmarking and Validation of \ds[\normalsize]}
\label{sec:validation}
\noindent
We now scrutinise our proposed \ds{} dataset, answering RQ4 (``\purple{\textit{what performance do previous methods achieve on \ds{}?}}'') in §\ref{ssec:perf_ds}; and RQ5 (``\purple{\textit{does \ds{} contain phishing emails of a higher quality than those included in previously-proposed datasets?}}'') in §\ref{ssec:user_study}.

\subsection{Performance Assessment {\normalsize [RQ4]}}
\label{ssec:perf_ds}

\noindent
We test the detectors considered in §\ref{sec:reassessment} on our \ds{} dataset. Importantly, we only consider the English portion of \ds{} (encompassing 11502 emails in total, 5996 phishing and 5506 benign ones). This is to enable a fair comparison with our previously evaluated methods, which are tailored for English texts.

\textbf{Method.} We follow the same evaluation protocol as in §\ref{ssec:setup}, but the major difference is that, here, we use \ds{} only to \textit{test} existing methods. This is to follow the guidelines of Arp et al.~\cite{arp2022and}, who recommend that the test data should chronologically follow the training data. Given that our \ds{} was generated in 2025, and that all our considered datasets contained data collected much earlier (see Table~\ref{tab:email_datasets}), it follows that our assessment resembles a realistic evaluation. This difference leads to two deviations: for Experiment-1, we simply test all of our (already trained/fine-tuned) models on \ds{}; for Experiment-2, we train/fine-tune each model on all our 8 considered datasets, and then test it on \ds{}. Finally, for Experiment-3, we expand the list of considered LLMs to cover 14 (up from 2) models, each tested on \ds{}.

\begin{figure}[htpb!]
\centering
\begin{subfigure}{0.8\linewidth}
\includegraphics[width=\textwidth]{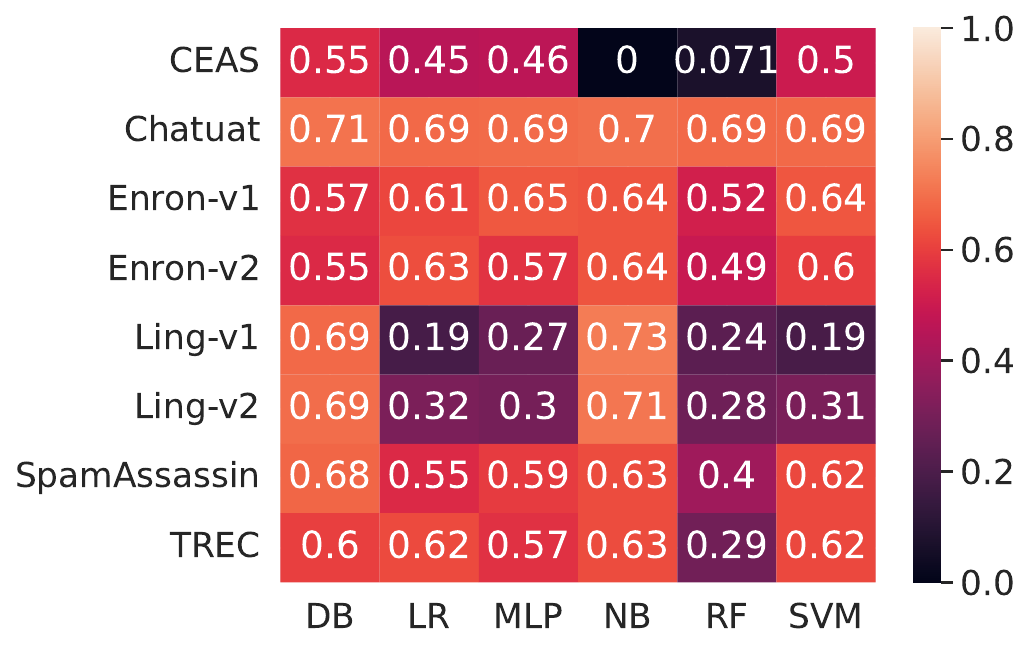}
\caption{Experiment-1 (cross-Evaluation) \textmd{English-only emails.}}
\label{fig:first}
\end{subfigure}
\hfill
\begin{subfigure}{0.6\linewidth}
\includegraphics[width=\textwidth]{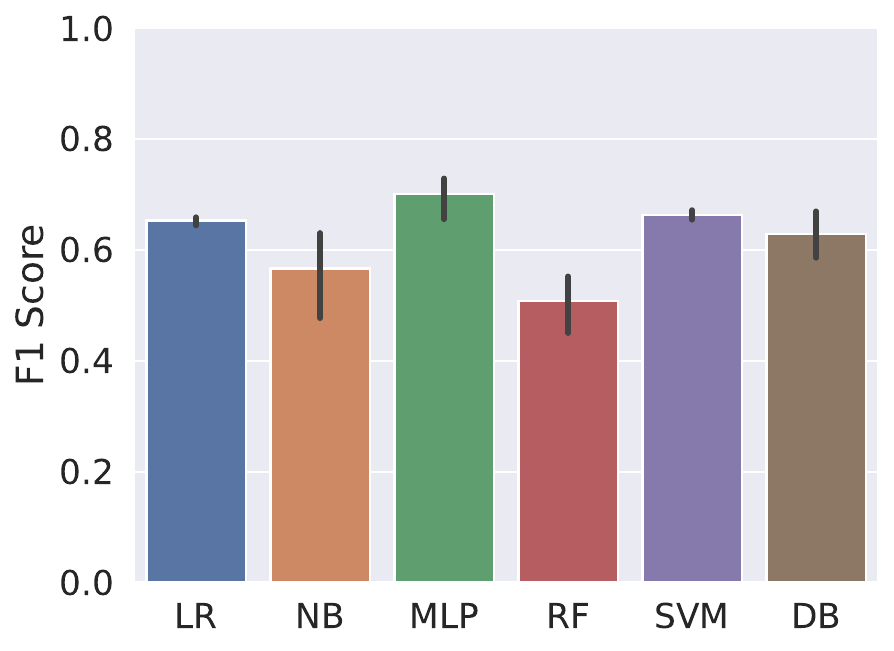}
\caption{Experiment-2 (all-vs-one). \textmd{English-only.}}
\label{fig:second}
\end{subfigure}
\hfill
\begin{subfigure}{0.8\linewidth}
\includegraphics[width=\textwidth]{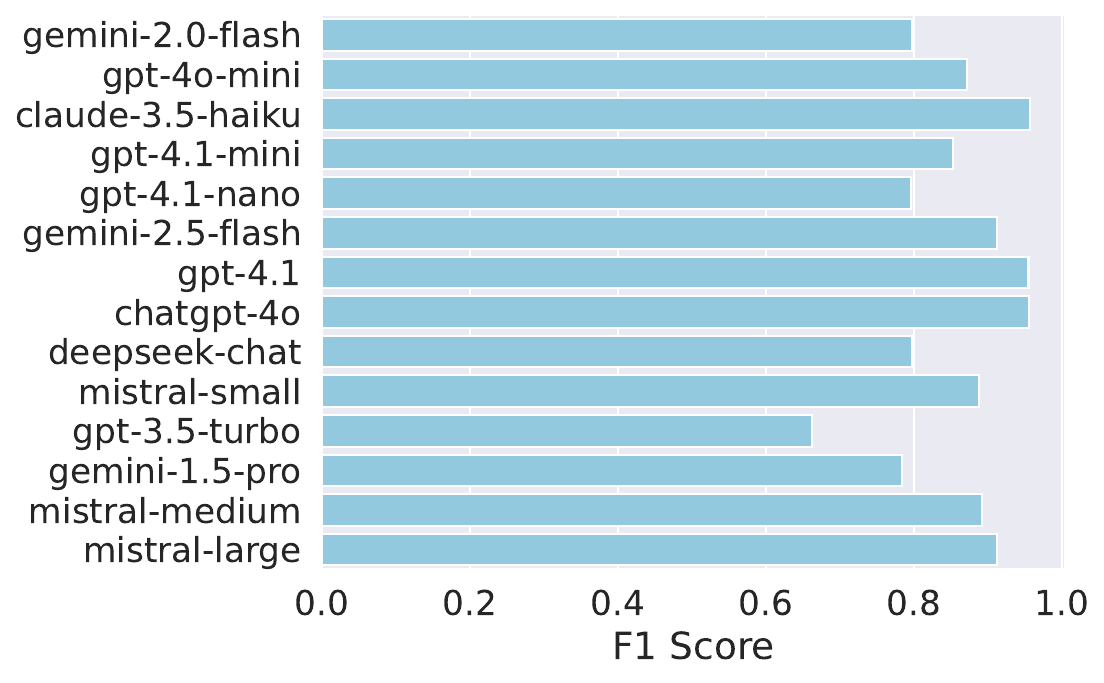}
\caption{Experiment-3 (LLMs). \textmd{English only emails.}}
\label{fig:third}
\end{subfigure}

\caption{Benchmarking existing detectors on our proposed \ds[\scriptsize] dataset. }
\label{fig:caripoti}
\end{figure}

\textbf{Results.}
We present the results of the experiments in Figure~\ref{fig:caripoti}.
We can highlight the following results:
\begin{itemize}[leftmargin=*]
    \item \textit{Experiment-1}: the cross-evaluation settings are very challenging for all of our models, with F1-score ranging from 0 (for the NB trained on CEAS) to 0.73 (for the NB trained on Ling-v1). The DB model has the most consistently-high F1-score (ranging 0.55--0.71), and the Chataut dataset also leads to the highest overall F1-score (ranging 0.69--0.71). 
    
    \item \textit{Experiment-2}: the all-vs-one setting shows a similar trend, i.e., the F1-score---while higher---does not go above 0.75. This indicates that our \ds{} dataset is substantially different from any of our eight considered datasets.
    
    \item \textit{Experiment-3}: intriguingly, LLMs show quite high and robust detection results, with 12 (out of 14) of the models exhibiting an F1-score \smamath{\geq}0.8. Some models, like \texttt{claude-3.5-haiku} shows brilliant performance, with an F1-score \smamath{\approx}0.95; the worst is gpt-3.5-turbo (F1-score \smamath{\approx}0.7). 
   
\end{itemize}
We report the complete results (standard deviations, as well as accuracy/precision/recall) in our repository~\cite{repository2}.


\begin{cooltextbox}
\textsc{\textbf{Answer to RQ4.}} Detectors trained on any combinations of our eight considered datasets struggle to detect the phishing emails in our \ds{} dataset. LLMs, however, are much more effective. These results show that our proposed \ds{} dataset represents a better ``benchmark'' than existing datasets to test previously proposed detectors.
\end{cooltextbox}

\subsection{User Study {\normalsize [RQ5]}}
\label{ssec:user_study}
\noindent
To fairly compare the quality of the emails included in \ds{} w.r.t. those included in existing datasets, we carried out a user study with 30 participants. To the best of our knowledge, we are the first to evaluate the quality of a phishing-email dataset in such a way.

\textbf{Procedure.}
We used convenience sampling~\cite{etikan2016comparison} to achieve a target number of 30 participants: each invitee was unaware of our research and had some background in cybersecurity. Each participant had to fill an online questionnaire, which began with some instructions and preliminary demographics questions. Then, each participant evaluated 20 phishing emails: 5 sampled from \ds{}, and the remaining 15 sampled from SpamAssassin, Enron, Nazario\footnote{We considered SpamAssassin and Enron because we also used these in our assessment (§\ref{sec:reassessment}); whereas we used Nazario as a representative of a popular dataset that has also more recent emails in it (see §\ref{ssec:datasets}).} (5 each). We created 10 versions of the questionnaire, so each set of 20 emails was seen by 3 participants---this enabled us to cover 200 emails in total (50 per dataset). To evaluate an email, the participant had to rate the ``overall phishing quality'' of the email, i.e., how convincing, well-written, and realistic it appeared; the rating was expressed via a 1--5 likert scale (1=low, 5=high). The questionnaires are available in our repository~\cite{repository2} (snippets are in Figures~\ref{fig:user-study-example}, \ref{fig:user-study-example2}).

\textbf{Sample Demographics and Results.}
Most participants were aged 18–25 (55\%) or 26–40 (45\%). Occupations included students (40\%) and various roles in tech, consulting, and other sectors. Regarding IT expertise, 52\% identified as intermediate, 22\% as expert, and 16\% as beginner. Cybersecurity experience was mostly intermediate (32\%), with others reporting beginner (22\%), no experience (23\%), advanced (20\%), or expert (3\%).
The average quality scores were: \ds{}=3.41 (std=1.2); Nazario=2.65 (std=1.3); SpamAssassin=1.57 (std=1.0); Enron=1.45 (std=0.9). Even a t-test confirms that the perceived quality of the emails from \ds{} is statistically superior to those of the other three datasets (\smamath{p}\smamath{<}\smamath{.05}).

\begin{cooltextbox}
\textsc{\textbf{Answer to RQ5.}} 
Yes, \ds{} contains phishing emails of higher quality than those of SpamAssassin, Nazario, and Enron.
\end{cooltextbox}

\section{Discussion}
\label{sec:discussion}

\noindent
Our open-problem paper should inspire a reflective exercise on research on phishing-email detection. Here, we state the limitations and ethics of our work, and then discuss the potential ``dual use'' of our proposed \fmw{}.

\subsection{Limitations and Disclaimers}
\label{ssec:limitations}
\noindent
We transparently outline the limitations of our paper.

First, we acknowledge that our literature review (in §\ref{ssec:datasets}) is not fully-systematic. Yet, this is not a problem because our conclusions (i.e., the fact that prior literature mostly focused on the same set of ``overly-used'' datasets) are shared also by prior work (see, e.g.,~\cite{das2019sok,alhuzali2025depth,salloum2022systematic,champa2024curated,kyaw2024systematic,thakur2023systematic}). Carrying out a systematic literature review is outside the scope of this open-problem paper.

Second, a threat to validity of our reassessment (§\ref{sec:reassessment}) is due to the potential labeling issues present in some of our considered datasets (in Table~\ref{tab:email_datasets}). We relied on the data provided by peer-reviewed prior work (including the Chataut dataset~\cite{chataut2024can}). We carried out manual checks and we believe that the labeling in this dataset is not foolproof. However, we refrain from changing the ground truth of previously-proposed datasets: we release all of our code (including our random seeds) so future work can reproduce our experiments on hypothetical ``fixed'' versions of our considered datasets.

Third, for our user study (§\ref{ssec:user_study}), we primed our users by mentioning that the study was about phishing: this may induce bias in the responses, but is a common practice in phishing studies~\cite{baki2023sixteen} and we refrained from using deception. Moreover, also related to our user study, we only considered a small sample of each of our considered datasets, and we only solicited the opinion of 30 participants---so we acknowledge that our investigation cannot cover all cases.

Finally, \textbf{we do not assert} that not releasing source code or using private dataset diminishes the contributions of prior work. Our reassessment is a way to strengthen prior work's contributions, since our intra-dataset results confirm the findings of prior literature.

\subsection{Ethical Considerations}
\label{ssec:ethics}

\noindent
Our institutions do not mandate that a formal IRB process is required to carry out the research discussed in this work. Still, we followed established ethical guidelines to carry out our research~\cite{bailey2012menlo}. No human was harmed as a result of our user study, we did not use any deception, and we explicitly asked for each subject's consent to participate in our study. We also gave our contacts so that participants could ask us to delete their responses, if they so desire. The questionnaire was anonymous, and participation was voluntary and we offered no compensation. With regards to our proposed \fmw{} framework, we acknowledge that parts of it can be exploited by malicious entities (e.g., to generate phishing emails). Yet, we believe that the risk is minimal: real attackers \textit{are well-aware} that LLMs can be used to craft phishing emails at scale~\cite{proofpoint2024phish}, and we believe that the publication of this paper (and of its resources) would not aggravate this risk.

\subsection{\fmw{} Dualism: Benign and Offensive AI}
\noindent
The \fmw{} exemplifies a dual-use technology with applications in both defense and offense.

\textbf{Benign Usage: Enhancing Cyber Defense.}  
\fmw{} can be used to improve phishing detection by generating synthetic data based on emerging phishing techniques. Cybersecurity experts can analyze novel attack patterns and then use the framework to create simulated phishing emails to fine-tune defense models. In addition, specialized data sets can be generated in various languages to improve the robustness of detection systems against diverse threats. The modularity of the framework allows companies to react to new threats quickly: if a novel phishing tactic emerges, companies can directly generate relevant company and employee profiles, define the characteristics of the new attack, and use the framework to simulate and incorporate these patterns into their internal detection systems, without the need to wait for real phishing attempts to be collected. This flexibility accelerates defense adaptation.

\textbf{Malicious Usage: Empowering Offensive AI.}  
Attackers can exploit OSINT (e.g., company websites, LinkedIn profiles) to gather detailed information about an organization. This data is fed into the framework’s first module to create accurate employee profiles. Using the second module, attackers can generate customized spear phishing emails that mimic organizational communication, lowering the expertise barrier for sophisticated attacks. Thus, \textbf{\fmw{}} serves as a tool for \textit{Offensive AI}~\cite{schracker2024sok}, making phishing attacks more accessible and effective.

\section{Recommendations and Future Work}
\label{ssec:conclusions}

\noindent
We showed that research on phishing-email detection is relatively stagnant due to the overuse of (outdated) benchmark datasets.

Our contributions serve as a scaffold to revitalize the domain of phishing-email detection. Our open-source reassessment, reliant on cross-evaluation experiments, can be used to create new testbeds by future work. Our proposed \fmw{} framework can facilitate the generation of novel datasets---which can augment our proposed \ds{} dataset. All such datasets, alongside being usable to test ``generic'' detectors of phishing emails, can also be used to devise ``specific'' detectors of LLM-generated emails---which is a subtle threat for which no solution exists.

We recommend future research to think deeply before using any given dataset for phishing-email detection to test their proposed methods. At the very least, the objective should be clearly stated: is it to ``outperform previously proposed methods'' or to ``develop a method that can detect phishing emails in the real-world''?

\begin{acks}
The authors thank the AISec reviewers for the great feedback. Parts of this research has been funded by the Hilti Foundation.    
\end{acks}

\bibliographystyle{ACM-Reference-Format}

\appendix
\section{User Study Snippets}
\begin{figure}[!hb]
    \centering
    \includegraphics[width=0.95\linewidth]{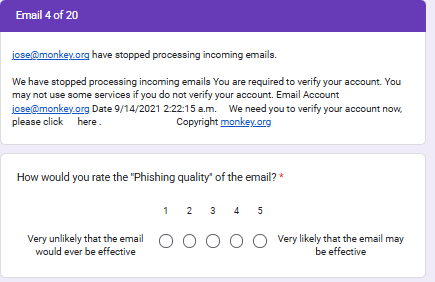}
    \vspace{-3mm}
    \caption{Illustrative example of an email evaluated in our user study. \textmd{The email is taken from the Nazario dataset.}}
    \label{fig:user-study-example}
\end{figure}

\begin{figure}[!hb]
    \centering
    \includegraphics[width=0.95\linewidth]{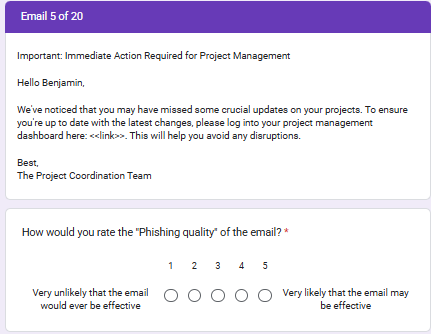}
    \vspace{-3mm}
    \caption{Illustrative example of an email evaluated in our user study. \textmd{The email is taken from the \ds{} dataset.}}
    \label{fig:user-study-example2}
\end{figure}

\section{Post-acceptance Responses (Q\&A)}
\label{app:post}
\noindent
The AISec reviewers provided great feedback. We would like to respond directly here to their suggestions/remarks, in a Q\&A format.

\textit{Would using different LLMs to generate dataset influence detection performance?}
It depends. There is evidence that LLMs have different writing styles (e.g.,~\cite{najjar2025leveraging}). Therefore, using a single LLM to generate a dataset would mean that any ML model trained on such a dataset would likely be effective at detecting emails generated by the same LLM that generated the dataset. Given that GPT-based models are widespread, we used these models to create \ds{}. However, we acknowledge that, for a truly comprehensive dataset that can cover the writing style of multiple LLMs, it would be desirable to expand our proposed \ds{} by generating additional samples using different LLMs (which we leave to future work). 

\textit{Could there be bias in the user study? Older datasets might reflect outdated phishing styles, while newer samples, shaped by current trends, could appear unfamiliar or more convincing simply because they differ from the ``already known'' types of phishing emails.} In our user study, we asked a specific question: ``How would you rate the `phishing quality' of the email?''. Our rationale is that, since our study is done in 2025, the rating provided by our participants would reflect the ``quality'' according to current trends. Note: it is implicit that, in claiming that \ds{} has emails of ``better quality'', \textit{we are specifically referring to the current state of phishing}. It would be false to state that the emails contained in previously proposed datasets to be of poor quality in the general sense (after all, those emails were \textit{real} phishing emails---but most such emails pertain to outdated phishing tactics, since they were exchanged, e.g., years before the advent of LLMs).

\begin{figure}[!tb]
    \centering
    \includegraphics[width=0.95\linewidth]{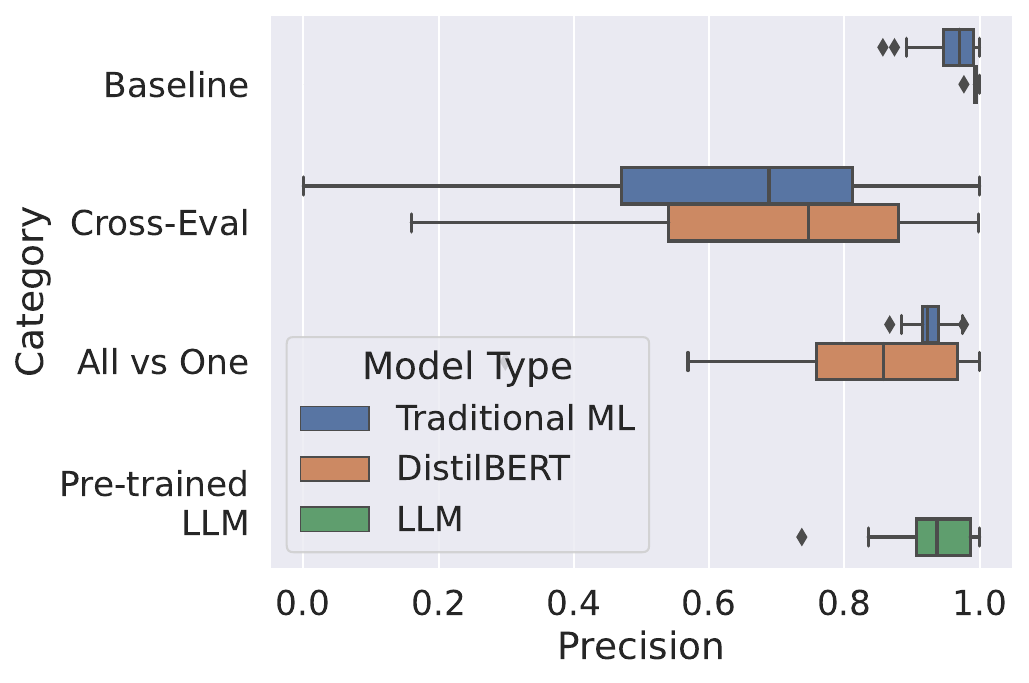}
    \vspace{-3mm}
    \caption{\textbf{Precision results of our reassessment experiments.} \textmd{Each boxplot represents the distribution of the precision ($\frac{TP}{TP+FP}$) across all of our experiments on existing datasets. For instance, the blue bins show the aggregated results of all the models using feature-based techniques.}}
    \label{fig:precision_boxplot}
    \vspace{-5mm}
\end{figure}

\textit{You discussed the F1-score in the paper, can you provide more insights on the false positives?} That's true. We focused on the F1-score because it is the only metric which accounts for both detection rate and false-positive rate. However, we agree that measuring false positives is a crucial metric for cyber-threat detection purposes. We provide additional metrics (recall, precision, accuracy) in our repository~\cite{repository2}. Moreover, and to get a broad overview of the effects of false positives, we report in Figure~\ref{fig:precision_boxplot} the distribution of the precisions (i.e., {\small $\frac{TP}{TP+FP}$}) obtained by aggregating the results of \textit{all} our models in the re-assessment experiments (in §\ref{sec:reassessment}); intuitively, a precision close to 1 indicates a low number of false positives. We see that the precision is very high in the baseline case (i.e., when a model is trained/fine-tuned and tested on data from the same ``source'' dataset), but it drops substantially when the test is done on a different dataset. Note that we do not fine-tune the LLMs, which is why they are placed in a different category.

\textit{Does dataset balancing matter?} The dataset we used in our reassessment, according to Table~\ref{tab:email_datasets}, have different compositions: some are balanced (e.g., Enron-v1) others are very unbalanced (e.g., Ling-v1). Yet, the same-dataset performance is always high even in highly-unbalanced cases. It is difficult to draw absolute conclusions in the cross-evaluation case, as performance varies even when the models are trained on imbalanced datasets. For instance, the MLP trained on Ling-v2 (highly unbalanced) has 0.53 F1-score on CEAS, but 0.79 on SpamAssassin; whereas the MLP trained on Enron-v2 (balanced) has 0.73 F1-score on CEAS, but 0.42 on SpamAssassin. We believe that, rather than ``balancing'', the major difference is the type of phishing email contained in each of these datasets.

\textit{The literature review does not include approaches from the Industry.} That is true. We did not cover these, because our literature review was rooted on academic work. (Recall the point of departure of our research was: ``what are the datasets commonly used in phishing-email-detection literature?'' Informally, we hope that, if ML is used in industry, the corresponding methods are trained on ``better'' datasets than those used in academic literature!) We therefore acknowledge that our analysis does not include approaches for industry. This could be a room for future work.

\textit{What about hyperparameter configurations?} We tried to align our reassessment to prior work. This is why we configured our models by using the parameters reported by prior work which have been found to work better in the specific context of phishing email detection. Given that our results align to those claimed by prior work, we believe our choice to be valid and functional to our purpose.

\textit{A valuable future direction would be to integrate E-PhishLLM-generated emails into a complete phishing campaign tool for controlled testing with real users. This would offer a realistic validation of the dataset’s effectiveness.} This is a wonderful suggestion that we will certainly consider pursuing.

\end{document}